\documentclass[letterpaper,twocolumn,10pt]{article}
\usepackage{usenix}

\usepackage{xspace}
\usepackage{spverbatim}

\usepackage{listings}
\usepackage{subcaption}
\usepackage{multirow}

\usepackage{afterpage}

\usepackage{tikz}
\usepackage{color}
\usepackage{xcolor}
\usepackage{textcomp}
\usepackage{listings}
\usepackage{pifont}
\usepackage{tcolorbox}
\usepackage{svg}
\usepackage{booktabs}
\usepackage{float}
\usepackage{amsmath}
\usepackage{hhline}
\usepackage{stmaryrd}
\usepackage{ulem}

\usepackage{nopageno}
\usepackage{xspace}
\usepackage{amsmath,amssymb,amsfonts}
\usepackage{algorithmic}
\usepackage{graphicx}
\usepackage{subcaption}
\usepackage{textcomp}
\usepackage{xcolor}
\usepackage{hyperref}
\usepackage{xurl}   
\usepackage{flafter}  
\usepackage{hhline}
\usepackage[title]{appendix}

\usepackage{pgfplots}                       \pgfplotsset{compat=1.18}

\usepackage{spverbatim}
\usepackage{listings}   
\usepackage{caption}    
\usepackage{multirow}

\usepackage{tabularx,soul,mdframed,multirow}

\usepackage{varwidth} 

\usepackage{algorithm}

\newcommand{\yes}{\ding{51}} 
\newcommand{\no}{\ding{55}}  
\PassOptionsToPackage{hyphens}{url}

\usepackage{color}
\definecolor{lightgray}{rgb}{.9,.9,.9}
\definecolor{darkgray}{rgb}{.4,.4,.4}
\definecolor{purple}{rgb}{0.65, 0.12, 0.82}

\lstdefinelanguage{JavaScript}{
  keywords={typeof, new, true, false, catch, function, return, null, catch, switch, var, if, in, for, while, do, else, case, break, instanceof},
  keywordstyle=\color{blue}\bfseries,
  ndkeywords={class, export, boolean, throw, implements, import, this},
  ndkeywordstyle=\color{darkgray}\bfseries,
  identifierstyle=\color{black},
  sensitive=false,
  comment=[l]{//},
  morecomment=[s]{/*}{*/},
  commentstyle=\color{purple}\ttfamily,
  stringstyle=\color{red}\ttfamily,
  morestring=[b]',
  morestring=[b]"
}

\def\systemname{\systemnameRaw\xspace}
\def\sys{\systemname}

\newcommand{\paragraphtitle}[1]{\vspace{5pt}\noindent\textbf{#1.}}

\newenvironment{icompacter}{
  \begin{list}{$\bullet$}{
    \parsep 0pt plus 1pt            
    \partopsep 0pt plus 1pt         
    \topsep 2pt plus 2pt minus 1pt  
    \itemsep 0pt plus 0pt           
    \parskip 2pt plus 1pt           
    \leftmargin 0.13in              
    \labelwidth 0.13in
    }}
  {\normalsize\end{list}}

\newcounter{circlednum}
\newcommand*\circled[1]{\tikz[baseline=(char.base)]{
            \node[shape=circle,fill,inner sep=0.5pt] (char) {\tiny \textcolor{white}{#1}};}}

\newcommand{\twodigitcircled}{%
  \ifnum\value{circlednum}<10 0\fi\arabic{circlednum}%
}

\usepackage{wasysym}

\definecolor{green}{rgb}{0.000, 0.506, 0.000} 
\definecolor{lightgreen}{rgb}{0.41, 0.66, 0.31}        
\definecolor{darkgray}{rgb}{0.4, 0.4, 0.4}        
\definecolor{purple}{rgb}{0.6, 0, 1}
\definecolor{pink}{rgb}{0.73, 0, 0.39}         
\definecolor{blue}{rgb}{0, 0, 1}                  
\definecolor{black}{rgb}{0, 0, 0}                 
\definecolor{regexcolor}{rgb}{0.73, 0.4, 0.53}    
\definecolor{builtincolor}{rgb}{0.90, 0.57, 0.22}        
\definecolor{stringcolor}{rgb}{0.61, 0.27, 0.22}  
\definecolor{pinkbg}{rgb}{1.0, 0.92, 0.95}
\definecolor{vueblue}{rgb}{0.168, 0.573, 0.686}

\lstdefinelanguage{JavaScript}{
  keywords={break, case, catch, continue, debugger, default, delete, do, else, false, finally, function, if, in, instanceof, new, null, return, switch, this, throw, true, try, typeof, var, void, while, with, const, Section, v-for, v-html, div, async, await},
  morecomment=[s]{/*}{*/},
  morestring=[b]',
  morestring=[b]",
  ndkeywords={class, export, boolean, throw, implements, import, this, URL, Error, URLSearchParams},
  keywordstyle=\color{blue}\bfseries,
  ndkeywordstyle=\color{green},
  identifierstyle=\color{black}\ttfamily,
  commentstyle=\color{purple}\ttfamily,
  stringstyle=\color{stringcolor}\ttfamily,
  sensitive=true,
  morekeywords=[3]{test, substring, getElementsByTagName, charCodeAt, push, setAttribute, getAttribute, apply, indexOf},
  keywordstyle=[3]\color{blue},
  morekeywords=[4]{scripts, length, textContent, nodeValue, nodeType},
  keywordstyle=[4]\color{blue}\ttfamily,
  escapechar=∫
}

\definecolor{codegreen}{rgb}{0,0.6,0}
\definecolor{codegray}{rgb}{0.5,0.5,0.5}
\definecolor{codepurple}{rgb}{0.58,0,0.82}
\definecolor{backcolour}{rgb}{0.95,0.95,0.92}

\lstdefinelanguage{MyPrompt}{
  keywords={System, Response, User},
  sensitive = true,
  comment=[l]{//}, 
  morecomment=[s]{/*}{*/}
}
\lstdefinestyle{mypromptstyle}{
    language=MyPrompt,  
    backgroundcolor=\color{white},
    basicstyle=\scriptsize\ttfamily,
    commentstyle=\color{codegreen},
    keywordstyle=\color{codepurple}\bfseries,
    numberstyle=\tiny\color{codegray},
    stringstyle=\color{blue},    
    breakatwhitespace=false,
    breaklines=true,
    captionpos=b,
    keepspaces=true,
    numbers=left,
    numbersep=5pt,
    tabsize=4,
    columns=fullflexible,
    morecomment=[s]{[}{]},
}

\newcommand{\shortvul}{\textsc{MPV}\xspace}

\newcommand{\shortvuls}{\textsc{MPVs}\xspace}

\newcommand{\vul}{{Message Progression Vulnerability}\xspace}

\newcommand{\vuls}{{Message Progression Vulnerabilities}\xspace}

\def\systemnameRaw{Proton}

\newcommand{\secref}[1]{\S\ref{#1}}

\newcommand{\phasetwo}{26\xspace}
\newcommand{\llmreportfp}{90\xspace}

\newcommand{\zeroday}{23\xspace}
\newcommand{\zerodayrce}{22\xspace}

\newcommand{\ack}{13\xspace}
\newcommand{\fix}{11\xspace}
\newcommand{\cve}{11\xspace}
\newcommand{\llmreport}{116\xspace}
\newcommand{\unflagged}{473\xspace}

\newcommand{\repourl}{https://anonymous.4open.science/r/proton\xspace}

\newcommand{\datasetsize}{589\xspace}

\begin{document}





%



%
%


\title{Buzz to Boom: Detecting Message Progression Vulnerabilities in Electron Applications via Segmented Directed Fuzzing}


\author{
Jianjia Yu, Zhengyu Liu, Ziyang Li, Yu Sun, and Yinzhi Cao \\
\{jyu122, zliu192, ziyang, ysun227, yinzhi.cao\}@jhu.edu \\
Johns Hopkins University
}


\maketitle

\begin{abstract}

Electron is a popular framework for building cross-platform desktop applications using web technologies.
Such applications consist of multiple processes with different privilege levels that communicate via message passing.
When inter-process messages carry attacker-controlled inputs, they can propagate across processes and reach privileged APIs, e.g., command execution.  Such  a message propagation behavior
 is characterized as Message Progression Vulnerabilities (MPVs).  The exploitation of MPVs is challenging because it often requires multiple steps, e.g., first arbitrary code execution in one process via message passing, and then command injection in another process using another message crafted in the first process. 
%
%
%
%
To our knowledge, existing works on Electron security only study unsafe configurations and malicious Document Object Model (DOM) content, i.e., they cannot detect or exploit these vulnerabilities that need to be triggered by complex cross-process exploits via message passing.

We present \sys, a segmented directed fuzzing framework for detecting MPVs.
Our key insight is to decompose end-to-end fuzzing into per-process segments along message-passing boundaries, where the goals of fuzzing each segment are either: (i) reaching a sink in the current process or (ii) propagating the payload to the next process, to enable the exploration of another process.
In the second case, the messages seed the corpus of the next segment.
Finally, \sys synthesizes crash inputs from each process to validate end-to-end exploits.
We evaluate \sys against \datasetsize real-world Electron applications, resulting in \zeroday zero-day MPVs. Among them, \zerodayrce lead to OS command execution, including projects with over 50k GitHub stars. We responsibly disclosed all findings. To date, we have received \ack acknowledgments, \fix fixes, and \cve CVEs, including a bug bounty from Vercel.

\end{abstract}

\begin{figure*}[htb]
    \includegraphics[width=\linewidth]{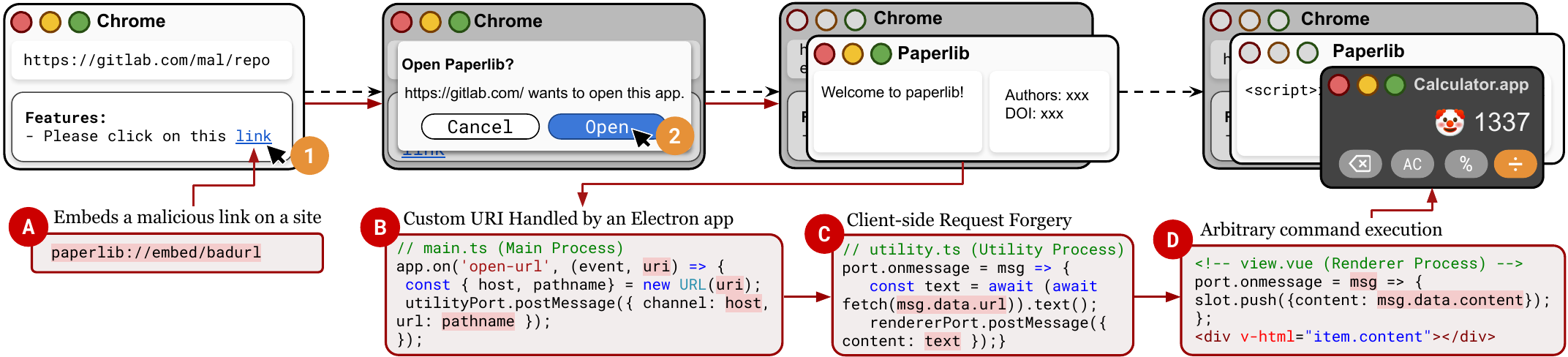}
    \vspace{-10px}
    \caption{
        An illustration of a zero-day \shortvul found by \sys in Paperlib: 
        a victim clicks on a malicious custom URI link \ding{192} embedded in a webpage, 
        which launches Paperlib after browser confirmation \ding{193}. 
    The URI payload progresses across three Electron processes, 
    main (B), utility (C), and renderer (D), via message passing, 
    ultimately yielding arbitrary command execution.
    }
    \label{fig:attack-workflow-victim}
\end{figure*}

\section{Introduction}
\label{sec:introduction}

Electron is a widely adopted framework for building cross-platform desktop applications using web technologies, powering popular applications such as Visual Studio Code, Slack, and Discord. 
Electron applications consist of browser-like renderer processes and a privileged Node.js-enabled main process that can access local files, spawn processes, and directly execute OS commands. These processes communicate through message-passing interfaces, e.g., Inter-Process Communication (IPC), as well as other cross-process operations including script execution, window navigation, and custom protocol handlers. 
When such cross-process channels carry attacker-controlled inputs, those inputs can progressively propagate across processes and reach privileged APIs, e.g., command execution.
We characterize such cross-process propagation of attacker-controlled data as \vuls (\shortvuls).

A noteworthy property of a \vul---differentiating it from existing ones~\cite{fass2021doublex, yu2023coco}---is the involvement of multiple Electron processes, leading to chaining two or more vulnerabilities through multi-step message passing, to the final consequence. Here is a three-step example. First, the message starts with a custom URI click in an external browser, and is then passed to the Electron main process for a client-side request forgery.  Second, the message progresses from the main process to the render in Electron applications to exploit an arbitrary code execution vulnerability in the render.  This still is not enough, though, because the render process is isolated without access to privileged APIs. Therefore, lastly, another message progression from the render back to the main process is needed to exploit another command injection vulnerability in the main.  The chaining of these three vulnerabilities consists of a \vul, whereas prior message-related vulnerabilities~\cite{fass2021doublex, yu2023coco,son2013postman, barth2008frame,steffens2020pmforce} only need at most one-step exploitation from one process to another.

Due to this property, the end-to-end detection and exploitation of \shortvul are challenging, and none of the prior works can achieve them.
%
%
%
%
%
First, existing Electron security studies mainly focus on unsafe configuration checks~\cite{doyensec2022electronegativity, ali2024rise, altpeter2020electron} and mitigation of malicious Document Object Model (DOM) content~\cite{yang2025coindef, jin2023security}, or unsafe cross-context communication~\cite{xiao2022understanding}.
 None of them detects the progression of attacker-controlled messages across multiple Electron processes, let alone \shortvul. 
%
Second, it is challenging to adapt existing techniques to detect \shortvul as well. 
%
%
 On the one hand, cross-context static analyses~\cite{fass2021doublex, yu2023coco} can identify suspicious message handlers or privileged API uses, but they struggle with the dynamic message construction and asynchronous message passing, which makes it difficult to determine whether a flagged path can be a feasible cross-process exploit chain or synthesize an exploit.
On the other hand, existing dynamic techniques such as web fuzzing~\cite{witcher, guler2024atropos, codeintelligence2023jazzerjs} are monolithic and do not coordinate across Electron's process boundaries.
 More specifically, inputs at each boundary are constrained by what survives the preceding boundary, i.e., existing fuzzers often waste their budget on inputs that fail to propagate past the first boundary.

In this paper, we present \sys, a segmented directed fuzzing framework for detecting and exploiting \vuls in Electron applications.
The key insight of \sys is to decompose the end-to-end fuzzing---potentially spanning over multiple processes---into independent segments along message-passing boundaries, to efficiently explore deeper program states.
%
%
%
%
%
%
More specifically, the fuzzing segmentation has two phases. Phase I performs static progression analysis to identify candidate vulnerable paths and segment boundaries by reasoning about payload injection points, IPC message flows, and dangerous API sinks.

Then, Phase II performs coverage-guided fuzzing on each segment independently.  There are two objectives: reaching a security-related, \textit{terminal sink} in the current process, or reaching an \textit{intermediate sink}, or so-called \textit{Progression API} that propagates the payload through messaging to another process. 
 When a fuzzing process on one segment successfully reaches a Progression API, the message it produces becomes the corpus for the next segment.
 \sys will then mutate the message content driven by a harness generated by Large Language Models (LLMs) to fuzz the next segment. 
 After all per-segment fuzzing, \sys synthesizes end-to-end exploits and validates them against the whole application.

We evaluate \sys on \datasetsize real-world Electron applications and discover \zeroday previously unknown vulnerabilities, including \zerodayrce that can be escalated to full RCE. The vulnerable applications include highly popular projects with over 50k GitHub stars. We responsibly disclosed all findings, receiving \ack acknowledgments, \fix fixes, and \cve CVE identifiers to date, including a bug bounty from Vercel for a vulnerability in Hyper. 
Compared to end-to-end fuzzing, \sys discovers 17 more zero-day vulnerabilities.


\paragraphtitle{Contributions} This paper makes three main contributions:

\begin{icompacter}

\item We characterize \vuls (\shortvuls) in Electron applications, where attacker-controlled inputs propagate across multiple processes via message passing, chaining two or more vulnerabilities through multi-step exploitation to ultimately reach privileged APIs such as command execution. 
%

\item We propose segmented directed fuzzing as a detection strategy for \shortvuls and realize it in \sys, a framework that decomposes multi-process exploit chains into segments at message-passing boundaries and fuzzes each segment toward either a local sink or a downstream-bound message. 

\item 
We evaluate \sys on \datasetsize real-world Electron applications, uncovering \zeroday zero-day \shortvuls, with \zerodayrce RCE exploits, and received \cve CVEs with \fix fixes and \ack acknowledgments, along with a bug bounty from Vercel. 
We open-source \sys to support future research.
\end{icompacter}

\section{\vul}
\label{sec:vulnerability}



In this section, we begin with the description of \vul in~\secref{sec:vuln-description}. Then, we present a systematization of Message Progression Vulnerabilities in~\secref{sec:vuln-sys}, including Electron's process model and execution contexts (~\secref{sec:process-model}), the payload sources (~\secref{sec:sources}), and sinks (~\secref{sec:sinks}). 
Finally, ~\secref{sec:threat-model} illustrates the threat model.

\subsection{Vulnerability Description}
\label{sec:vuln-description}

%
%


Figure~\ref{fig:attack-workflow-victim} illustrates a zero-day \shortvul in Paperlib: a victim clicks a malicious \texttt{paperlib://} URI embedded in a webpage (box A), which triggers a chain of three vulnerabilities across Electron's processes via message passing — an unintended function invocation in the main process (box B), a client-side request forgery in the utility process (box C), and DOM-based XSS in the renderer (box D) — ultimately yielding arbitrary command execution.
This four-step chain, spanning three Electron processes, is a \shortvul. The chaining of vulnerabilities through message progression turns a browser click into full RCE. We present the vulnerability in detail in Section~\ref{sec:motivating}. 
To \shortvul generalize beyond this example, we next describe the process model and execution context of Electron applications, and then systematize \shortvul's sources and sinks.

\subsection{Vulnerability Systematization}
\label{sec:vuln-sys}

\begin{table*}[t]
\centering
\caption{Categorization of \vul sinks across Electron's multi-process architecture. We distinguish them by their effects: \textit{intermediate sinks} ({$\nearrow$}) forward attacker-controlled data into a message destined for another process; \textit{terminal sinks} ({$\square$}) directly cause the security consequence (e.g., OS command execution); a single API may act as either depending on the runtime context ({$\nearrow$ / $\square$}). For each sink type, we show: its effect, its availability across processes and execution contexts, potential payload sources, vulnerable code patterns, and exploitation conditions, where \texttt{T} denotes tainted data.}
\label{tab:sinks}
\scriptsize
\setlength{\tabcolsep}{2pt}
\begin{tabular}{@{}ll@{\hspace{8pt}}ccccc@{\hspace{8pt}}ll@{}}
\toprule
\multirow{4}{*}[-0.5em]{\textbf{Description}} & 
\multirow{4}{*}[-0.5em]{\textbf{Effect}} & 
\multicolumn{4}{c}{\textbf{API Location}} & 
\multirow{4}{*}[-0.5em]{\shortstack{\textbf{Payload}\\\textbf{(U/F/S)}}} &
\multirow{4}{*}[-0.5em]{\textbf{Code Patterns} } & 
\multirow{4}{*}[-0.5em]{\textbf{Conditions}} \\
\cmidrule(lr){3-6}
& & \multirow{2}{*}[-0.5em]
{\shortstack{\textbf{Main}\\\textbf{Proc.}}} & 
\multirow{2}{*}[-0.5em]{\shortstack{\textbf{Util.}\\\textbf{Proc.}}} & 
\multicolumn{2}{c}{\shortstack{\textbf{Renderer Proc.}}} & 
& & \\
\cmidrule(lr){5-6}
& & & & \shortstack{\textbf{Preload}} & \shortstack{\textbf{Main}} & & & \\
\midrule

\multirow{2}{*}{Inter-Process Communication} & \multirow{2}{*}{$\nearrow$} & \multirow{2}{*}{\CIRCLE} & \multirow{2}{*}{\CIRCLE}  & \multirow{2}{*}{\CIRCLE} & \multirow{2}{*}{\CIRCLE} & \multirow{2}{*}{U/F/S} & \texttt{ipcRenderer.send(C, T)} & \multirow{2}{*}{\texttt{T} contains attack-controlled payload} \\ 
& & & & & & & \texttt{port.postMessage(T)} \\
\midrule

\multirow{2}{*}{Remote Source Fetching} & \multirow{2}{*}{$\nearrow$} & \multirow{2}{*}{\CIRCLE} & \multirow{2}{*}{\CIRCLE}  & \multirow{2}{*}{\CIRCLE} & \multirow{2}{*}{\CIRCLE} & \multirow{2}{*}{U/F/S} & \texttt{fetch(T)} & \multirow{2}{*}{\texttt{T} contains attack-controlled payload} \\ 
& & & & & & & \texttt{https.get(T)} \\
\midrule

\multirow{3}{*}{Window Navigation} & 
\multirow{3}{*}{$\nearrow$ / $\square$} & 
\CIRCLE & \Circle & \Circle & \Circle & 
U/(F)$^2$/(S)$^3$ & 
{\tt win.loadURL(T)} & 
\texttt{T} is URL pointing to attacker-controlled resource \\
& & \CIRCLE & \Circle & \Circle & \Circle & 
U/F & 
{\tt contents.loadFile(T)} & 
\texttt{T} is path pointing to attacker-controlled local file \\
& & \Circle & \Circle & \CIRCLE & \CIRCLE & 
U/(F)$^2$/S & 
{\tt window.location.href=T} & 
\texttt{T} is URL pointing to attacker-controlled resource \\
\midrule

Subview Embedding & $\nearrow$ / $\square$ & \Circle & \Circle & \CIRCLE & \CIRCLE & 
U/(F)$^2$/S & 
{\tt iframe.src=T; webview.src=T} & 
\texttt{T} is URL pointing to attacker-controlled resource \\
\midrule

\multirow{2}{*}{Window Script Execution} &  \multirow{2}{*}{$\nearrow$ / $\square$} & 
\CIRCLE & \Circle & \Circle & \Circle & 
U & 
{\tt contents.executeJavaScript(T)} & 
\texttt{T} contains attacker-controlled JavaScript code \\
& & \Circle & \Circle & \CIRCLE & \Circle & 
U & 
{\tt webFrame.executeJavaScript(T)} & 
\texttt{T} contains attacker-controlled JavaScript code \\
\midrule

DOM-based Script Exec. & $\nearrow$ / $\square$
& \Circle & \Circle & \CIRCLE & \CIRCLE & 
U & 
{\tt element.innerHTML=T} & 
\texttt{T} contains attacker-controlled HTML/JS code \\
\midrule

Code Execution & $\nearrow$ / $\square$  & \CIRCLE & \CIRCLE & \CIRCLE & \CIRCLE & 
U & 
{\tt new Function(T); eval(T)} & 
\texttt{T} contains attacker-controlled JS code \\
\midrule

Command Execution & $\square$ &  \CIRCLE & \CIRCLE & \LEFTcircle & \LEFTcircle & 
U & 
{\tt exec(T); spawn(T, \{shell:true\})} & 
\texttt{T} contains attacker-controlled shell commands \\
\midrule

Module Loading  & $\square$  & \CIRCLE & \CIRCLE & \LEFTcircle & \LEFTcircle & 
U/(F)$^1$ & 
{\tt require(T); import(T)} & 
\texttt{T} is path to attacker-controlled local JS file \\
\midrule

\multirow{2}{*}{External Open Service} & 
\multirow{2}{*}{$\square$} &
\CIRCLE & \Circle & \LEFTcircle & \LEFTcircle & 
U/S & 
{\tt shell.openExternal(T)} & 
\texttt{T} is URL to attacker-controlled resource \\
& & \CIRCLE & \Circle & \LEFTcircle & \LEFTcircle & 
U/F & 
{\tt shell.openPath(T)} & 
\texttt{T} is path to attacker-controlled local file \\
\midrule

File System Write & $\square$ & \CIRCLE & \CIRCLE & \LEFTcircle & \LEFTcircle & 
U & 
{\tt fs.writeFile(T$_1$, T$_2$)} & 
\texttt{T$_1$} (path) and \texttt{T$_2$} (content) are attacker-controlled \\

\bottomrule
\end{tabular}

\begin{flushleft}
\textbf{Effect:} $\nearrow$: Intermediate sink, $\square$: Terminal sink;
\textbf{API Location:} \CIRCLE\xspace: Sink available under default secure configuration (\texttt{nodeIntegration=false}, \texttt{contextIsolation=true}, and \texttt{sandbox=true}); \LEFTcircle\xspace: Sink available under insecure configuration (\texttt{nodeIntegration=true}, \texttt{contextIsolation=false}, or \texttt{sandbox=false}); \Circle\xspace: Sink not available under any configuration. \textbf{Payload:} U=Custom URI, S=Remote Server Response, F=Local File Access; parentheses indicate optional. $^{1}$\xspace: Local file not necessary for \texttt{import} as it supports \texttt{data:} protocol. $^{2}$\xspace: File protocol may be necessary to bypass CSP. $^{3}$\xspace: Remote server not necessary when \texttt{T} uses \texttt{javascript:} or \texttt{data:} protocol.
\end{flushleft}
\end{table*}

\subsubsection{Process Model and Execution Context}
\label{sec:process-model}
%
Electron employs a multi-process architecture consisting of a privileged main process and multiple utility and renderer processes. 
The main process runs in Node.js with full native API access, controlling the application lifecycle.
Utility processes are spawned by the main to handle background tasks such as network requests and file access, and also run with Node.js access.
Renderer processes display web content and interact with users. Furthermore, to facilitate communication with the main process, Electron provides a group of privileged APIs through preload scripts configured per window. Under the default secure configuration, each renderer maintains two isolated JavaScript contexts: (1) the \textit{preload context} executes with access to privileged APIs (e.g., \texttt{process} and \texttt{ipcRenderer}) and selectively exposes functionality to web content via \texttt{contextBridge}; (2) the \textit{main world context} executes web page code with access only to standard browser APIs and the APIs exposed by preload scripts. This isolation prevents web content from directly accessing privileged APIs while enabling main-process services via IPC.

The different processes and execution contexts with distinct privilege levels define the boundaries across which message progression happens, and determine which inputs can be potential payload sources and whether an API is a progression API or a terminal sink.
We next categorize sources and sinks along these dimensions in~\secref{sec:sources} and ~\secref{sec:sinks}.


\subsubsection{Sources} 
\label{sec:sources}

Attacker-controlled payloads that originate outside the application's trust boundary can enter an Electron application via OS-level handlers or application logic. We identify three such source types, as shown in Table~\ref{tab:sinks}.

\begin{icompacter}
\item \textbf{Custom URI (U).} The attacker embeds a malicious payload within a URI using the application's registered custom scheme, e.g., \texttt{app://action?content=payload}. When a victim clicks the link in a browser, such as Chrome, the OS invokes the Electron application and delivers the URI to its \texttt{open-url} handler, where attacker-controlled parameters enter the application and may propagate to downstream sinks via IPC and other message progression channels. 

\item \textbf{Remote Server Response (S).} 
The application fetches content from an attacker-controlled server when attacker-controlled values flow to network APIs such as \texttt{fetch}.
%
In \shortvul chains, this source commonly arises when an intermediate vulnerability, such as a client-side request forgery (CSRF), happens. 
For example, the initial custom URI \texttt{app://install?url=evil.com} causes the application to fetch and process attacker-controlled content from \texttt{evil.com} through CSRF, and the server response is further processed. This pattern commonly appears in applications supporting plugin installation, remote file loading, and MCP server imports.

\item \textbf{Local File Access (F).} The application loads content from the local file system at an attacker-specified path, when attacker-controlled values flow to file system APIs (e.g., \texttt{fs.readFile}) or BrowserWindow methods (e.g., \texttt{win.loadFile}). No prior file system compromise is required: an attacker can exploit browser auto-download to silently place a malicious file in the default downloads directory, then trigger a custom URI to cause the application to load it, e.g., \texttt{app://open?path=\textasciitilde/Downloads/evil.js}.
\end{icompacter}

\subsubsection{Sinks}
\label{sec:sinks}

In Table~\ref{tab:sinks}, we list the \shortvul sinks across Electron's multiple processes and execution contexts.
We distinguish them by their effects: \textit{intermediate sinks} forward attacker-controlled data into a message destined for another process; \textit{terminal sinks} directly cause the security consequence (e.g., OS command execution).
Note that some APIs act as either intermediate or terminal sinks depending on the process and the execution context — e.g., whether it has Node.js access, as listed at lines 3-7. 
In the following paragraphs, we classify each sink by its primary role and note conditional behavior where applicable.

\vspace{-0.5em}

\paragraph{Intermediate Sinks.} Intermediate sinks are the APIs that forward attacker-controlled content to another process, toward a terminal sink.
They define the segment boundaries at which \sys decomposes exploit chains for per-segment fuzzing, and also serve as one of the two subjects of segment fuzzing. 
Specifically, we identify two types of intermediate sinks, the IPC-related APIs and the window navigation/embedding APIs.

First, IPC communications. These are the most prevalent intermediate sink in our dataset.
The most common patterns are \texttt{ipcRenderer.send(C, Msg)} and \texttt{port.\allowbreak postMessage(Msg)}, where \texttt{C} is the channel field that routes the message, and \texttt{Msg} is the content field. 
A summary of all types of IPC communication mechanisms in Electron applications can be found in Table~\ref{tab:ipc} in the Appendix.

Second, navigation and embedding APIs. These cause a process to load new content into an execution context, implicitly forwarding attacker-controlled URLs or paths, such as \texttt{win.loadURL(T)} and \texttt{iframe.src=T}.
These APIs act as intermediate sinks when the loaded content introduces further sinks in a new execution context. They become terminal when the loaded resource directly causes a security consequence. For example, when \texttt{T} starts with \texttt{javascript:}, so the code is directly executed, or \texttt{T} is a \texttt{file://} path to a local HTML file and Node is enabled, in which case the file can embed JavaScript code calling Node APIs.

\vspace{-0.5em}
\paragraph{Terminal Sinks}
%
Terminal sinks directly cause the final security consequence without further message progression, including OS command execution, module loading, and external service invocation. 
Beyond direct code execution sinks, we also consider arbitrary file write operations, as they can be readily escalated to code execution by overwriting application source code or system configuration files (e.g., \texttt{.bashrc} and desktop autostart entries) that execute during application or system startup.
%

The one type that is conditionally terminal are the code execution APIs, as shown in lines 4-6 in Table~\ref{tab:sinks}. APIs such as \texttt{contents.\allowbreak executeJavaScript(T)} and \texttt{eval(T)} evaluate attacker-controlled JavaScript. 
When reached in a process with Node.js access, i.e., main, utility, or a renderer with Node enabled, these APIs yield remote code execution directly and act as terminal sinks. 
When reached in a sandboxed renderer without Node.js access, they instead act as intermediate sinks. The attacker uses the gained JavaScript execution to invoke IPC APIs exposed via the preload script, progressing the payload to a privileged process where a terminal sink can be reached.

\begin{figure*}[ht]
    \includegraphics[width=\linewidth]{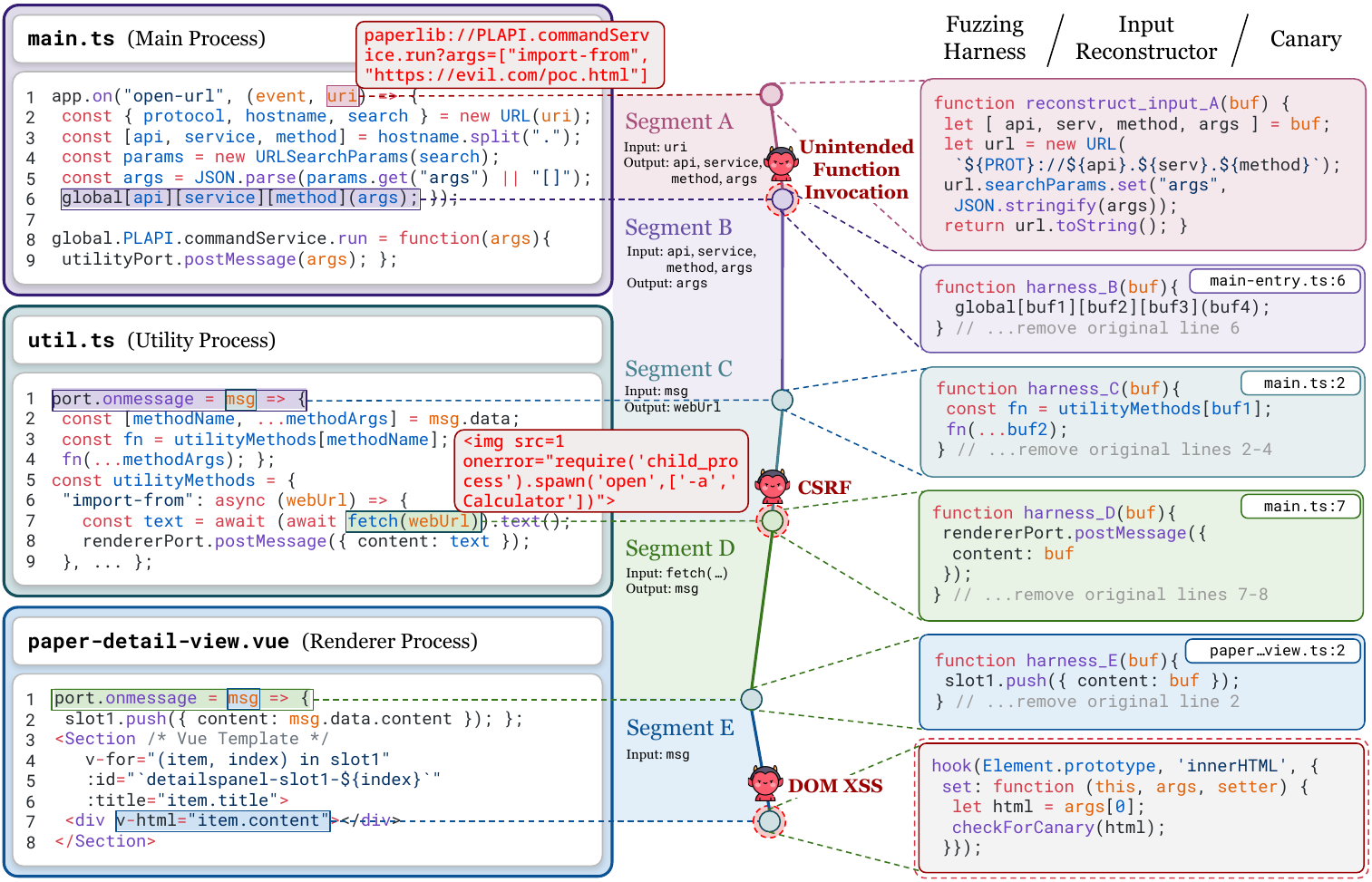}
    \vspace{-10px}
    \caption{
        Exploitable control-flow path in Paperlib~\cite{paperlib2025} discovered by \sys: 
        code snippets (left) from files running in the main, utility, and renderer processes; 
        agentic analysis segments the control flow into five segments A–E (middle); 
        LLM-synthesized artifacts on the right include an input reconstructor (A), fuzzing harnesses (B–E), and runtime canaries for final exploit validation (E).
    }
    \label{fig:motivating}
\end{figure*}

\subsection{Threat Model}
\label{sec:threat-model}

We consider an end user with a vulnerable Electron desktop application that has registered a custom URI scheme with the OS. Unlike prior works~\cite{ali2024rise, jin2023security, xiao2022understanding, yang2025coindef} that assume attackers already control renderer content, we consider a remote web attacker operating entirely outside the application, who can publish or embed crafted URIs on third-party sites or their own webpage without injecting content into the app directly. When the victim clicks a crafted link or visits an attacker-controlled page in a standalone browser, the OS invokes the target Electron application to handle the URI, leading to arbitrary command execution. We assume the application need not be already running, as the protocol is pre-registered with the OS, and no further user interaction is required.

\section{Overview}
\label{sec:overview}

We start with a motivating example in Section~\secref{sec:motivating}, then describe the key detection challenges and present an overview of our segmented fuzzing solution in Section~\secref{sec:challenge-solution}.


\subsection{A Motivating Example}
\label{sec:motivating}

Figure~\ref{fig:motivating} presents a real-world zero-day \vul (CVE-2025-XXXX) that leads to remote code execution, discovered by \sys in Paperlib~\cite{paperlib2025}, a popular open-source reference management system. 
The vulnerability lies in its core functionality that is designed to enable import of reference lists from external resources via \texttt{paperlib://} URI scheme.
The exploit requires chaining several vulnerabilities.
By crafting a malicious URI, an attacker first causes an unintended privileged function invocation, which then leads to a client-side request forgery (CSRF), triggering the application to fetch attacker-controlled content. Finally, when the content is processed, the payload exploits a DOM cross-site scripting to achieve arbitrary OS command execution.
We have responsibly disclosed this vulnerability to the developers, who have acknowledged and patched it.

\paragraphtitle{Vulnerability Details}
The application registers a custom URI handler for Electron's \texttt{open-url} event in the main process (\texttt{main.ts}, line 1). When a user clicks a malicious URI, the handler parses the URI into components (line 2) using \texttt{new URL}, extracts the API path by splitting the \texttt{hostname} (line 3), and retrieves arguments from the query parameters (lines 4-5). 
The handler then uses bracket notation to dynamically resolve and invoke the specified method on the global object (line 6), which is not expected to be accessible from custom URIs  — constituting an \textit{unintended function invocation}. 
In this case, the attacker's URI invokes \texttt{global[api]\allowbreak[service]\allowbreak[method]\allowbreak(args)} with malicious arguments.
The resolved method \texttt{commandService.run} (line 8) forwards these attacker-controlled arguments to the utility process via \texttt{postMessage} (line 9).

Next, the utility process (\texttt{util.ts}) receives the message through its \texttt{onmessage} handler and destructures the method name and arguments (lines 1-2), resolves the corresponding function from the \texttt{utilityMethods} object, and invokes it (line 3-4). The \texttt{import-from} method (lines 5-9) receives the malicious URL as its argument, fetches content from the attacker-controlled server using \texttt{fetch} (line 7)— a \textit{client-side request forgery} that introduces an additional remote server response payload (Source S).
The utility process then forwards the fetched text to the renderer process via \texttt{postMessage}. 

Finally, the renderer process (\texttt{paper\allowbreak-detail\allowbreak-view.vue}) receives the fetched content through \texttt{onmessage} (line 1) and pushes it into the \texttt{slot1} array (line 2). The Vue template (lines 3-7) then iterates over \texttt{slot1} and renders each item's content using the \texttt{v-html} directive (line 7), interpreting the content as raw HTML and executes any embedded JavaScript, constituting a \textit{DOM-based XSS} sink.

\paragraphtitle{Exploitation}
Since the renderer process is configured with node integration enabled and context isolation disabled, JavaScript executed in the renderer gains direct access to Node.js APIs. 
To host Source S, we publish a malicious PDF on a public file hosting service.
In our proof-of-concept, the injected script achieves arbitrary command execution via \texttt{require('child\_process').spawn()}, launching \texttt{Calculator.app} on the victim's machine.

\subsection{Challenges and Our Solutions}
\label{sec:challenge-solution}


The key challenge in detecting \shortvuls lies in their chained nature: exploit chains span multiple processes and sources via message progression, where several vulnerabilities compose into a severe consequence. Using the Paperlib example in Figure~\ref{fig:motivating}, we describe three concrete detection challenges this poses, and present how \sys addresses them through segmented fuzzing.

\paragraphtitle{Challenge I: Multiple payload sources}
Detecting \vul requires generating inputs from multiple heterogeneous sources that are introduced at different points in the execution flow.
In the motivating example, detecting the vulnerability in Paperlib requires generating two separate payloads introduced at different places: the custom URI (Source U) processed by the main process, and the remote server content (Source S) fetched by the utility process.
Furthermore, S depends causally on U. The URI determines which utility method executes, which then fetches the malicious content from the attacker's server. 
Traditional fuzzing cannot efficiently handle this because it operates with a single input source at program entry. It has no mechanism to discover S, which only appears after U reaches \texttt{fetch()}, nor to fuzz each source independently rather than exploring their cross-product.
%

\paragraphtitle{Challenge II: Customized and complex parsing}
Beyond multiple sources, each input must satisfy application-specific parsing constraints that vary across applications.
In Paperlib, the URI hostname is split into three components (main process, line 3) that resolve to a valid method path like \texttt{global["PLAPI"]["commandService"]\allowbreak ["run"]} (line 6). The \texttt{args} parameter must be valid JSON (line 5) with the correct method name as the first array element to trigger the fetch operation (utility process, line 7). The server response must be valid HTML text containing JavaScript that exploits the \texttt{v-html} sink (renderer process, line 7).
Traditional fuzzing cannot efficiently generate such inputs due to the sparse valid input space, while symbolic execution struggles to model complex parsing operations like JSON and URL parsing.

\paragraphtitle{Challenge III: Cross-process Execution Overhead} 
The multi-process architecture of Electron makes deep state exploration prohibitively expensive due to multiplicative cross-process execution overhead.
In the Paperlib example, exploring different states in the renderer process (lines 3-7 in \texttt{paper-detail-view.vue}) requires repeatedly executing the entire cross-process chain starting from the main process through the utility process. 
As chains grow longer across more processes, this multiplicative cost drastically reduces fuzzing throughput, limiting the ability to explore and generate inputs for deep program states where vulnerabilities may reside.

\paragraphtitle{Our Solutions}
\sys addresses all three challenges through \textit{segmented fuzzing}, that is, decomposing the complete exploitable control-flow path at Progression APIs and invertible parsing operations, then fuzzing each segment separately. 
As shown in Figure~\ref{fig:motivating}, \sys first performs agentic static analysis to identify a candidate vulnerable path and its segment boundaries, then generates fuzzing harnesses (B--E) and an input reconstructor (A) for each segment.
It then conducts segmented fuzzing to verify their reachability and generate end-to-end proof-of-concept inputs.
We now explain how this addresses each challenge.

First, \sys defines operations where attacker-controlled input crosses into a new execution surface as \textit{Progression APIs}, and segments the exploit path at these boundaries. 
In the Paperlib example, \texttt{fetch()} is a Progression API, as it receives the attacker-controlled URL from Source U and forwards it to an external server whose response becomes Source S. 
By segmenting at \texttt{fetch()}, \sys fuzzes the upstream segment to discover whether the attacker can control the request URL, then focuses on fuzzing the downstream segment over the server response independently.

Second, \sys handles complex parsing operations by leveraging their invertibility. Operations like \texttt{JSON.parse()} and \texttt{new URL()} have the property that given any valid output, a corresponding input can be reconstructed. \sys exploits this through the input reconstructor (Segment A in Figure~\ref{fig:motivating}): instead of generating raw URI strings that mostly fail parsing, the reconstructor works backward from valid parsed outputs to produce well-formed inputs that satisfy parsing constraints directly.

Third, \sys fuzzes each segment separately with per-segment harnesses, each of which initializes the segment's execution state directly without replaying preceding segments. For instance, harness E fuzzes \texttt{content} directly, bypassing main process URI parsing, utility process fetch, and all IPC communication, allowing \sys to explore the renderer at native fuzzing speed rather than replaying the entire chain per iteration. 
Once a valid segment chain is found, \sys sythinsizes the complete payload and validates end-to-end exploitability against the whole application.

\section{Methodology}
\label{sec:design}

\sys employs a two-phase methodology to efficiently uncover \shortvuls in Electron applications. 
First, an agentic static analysis phase identifies source–sink paths and segmentation boundaries, and generates per-segment harnesses. 
Second, a segmented fuzzing phase exercises each segment independently using generated harnesses. 
Finally, \sys performs end-to-end validation to compose the segment-level inputs into a full exploit payload and validates the path. 
We begin with a complexity analysis (\secref{subsec:insight}), then describe agentic static analysis (\secref{subsec:static}), segmented fuzzing (\secref{subsec:fuzzing}), and end-to-end validation (\secref{subsec:validation}).

\subsection{Segmentation Strategies}
\label{subsec:insight}

We first define Progression APIs, based on which \sys identifies segment boundaries in the application code. We then discuss how segmenting at these boundaries reduces the complexity of fuzzing.

\paragraphtitle{Progression APIs}
We define a \textit{Progression API} as an API call that, when reached with attacker-controlled input, forwards that input to a new execution surface, enabling a new independently-fuzzable segment to begin. 
Progression APIs generalize the notion of message passing beyond explicit IPC, as listed in Table~\ref{tab:sinks} at lines 1-5, along with their availability across processes and execution contexts. 
They include IPC mechanisms (\texttt{ipcRenderer.send}, \texttt{port.postMessage}), 
external fetch operations (\texttt{fetch(T)}) whose responses introduce new attacker-controlled content, 
and navigation and embedding APIs (\texttt{win.loadURL(T)}, \texttt{iframe.src=T}) that load attacker-controlled content into a new execution context. 
Additionally, code execution APIs (\texttt{eval(T)}, \texttt{executeJavaScript(T)}) act as Progression APIs when reached in a sandboxed renderer without Node.js access, where the gained execution enables IPC escalation toward a privileged process. 
\sys identifies Progression APIs as segment boundaries, and equivalently, \textit{intermediate sinks} from the fuzzing perspective. The goals of fuzzing each segment are either a Progression API, which seeds the next segment's corpus with its output, or a sink that directly manifests the vulnerability.

\paragraphtitle{Problem Definition}
The goal of \sys is to identify concrete exploitable control-flow paths where attacker-controlled inputs flow from one or multiple sources to a terminal sink, and decompose each path into independently-fuzzable segments at intermediate sinks, or Progression APIs. 
For any candidate path $\pi$, \sys decomposes it into consecutive segments $\text{seg}(\pi) = \{S_1, S_2, \ldots, S_k\}$, where each segment $S_i$ is a program fragment with:
(i) an input from a new external source (custom URI, server response, or file) or the output of segment $S_{i-1}$, and
(ii) an output that either reaches a Progression API to forward to $S_{i+1}$, or a terminal sink in $S_k$.

\begin{table}[t]
\centering
\caption{Instruction complexity and speedup comparison across segmentation strategies.}
\label{tab:complexity}
\footnotesize
\setlength{\tabcolsep}{9.5pt}
\renewcommand{\arraystretch}{2}
\begin{tabular}{@{}llcl@{}}
\toprule
\textbf{Strategy} & \textbf{\#Instructions} & \textbf{Speedup ($\times$)} & \textbf{Example} \\
\midrule
Baseline & $O((N_a+N_b)|I_a||I_b|)$ & -- & End-to-end \\
\midrule
DF Indep. & $O(N_a|I_a|+N_b|I_b|)$ & $\dfrac{|I_a||I_b|}{|I_a|+|I_b|}$ & Multi-source \\[0.3em]
Partial Dep. & $O(N_a|I_a| + N_b|I_a'||I_b|)$ & $\dfrac{|I_b|}{|I_a'|}$ & IPC \\[0.3em]
Fully Dep. & $O(N_b|I'_a|)$ & $\dfrac{|I_a|}{|I'_a|}$ & URL parsing \\[0.3em]
\bottomrule
\end{tabular}
\begin{flushleft}
\scriptsize
\end{flushleft}
\end{table}



\paragraphtitle{Complexity of an End-to-End Fuzzing Campaign}
We characterize the cost of a fuzzing campaign by the size of its \textit{program state space}, which we upper-bound by the product of:
(i) the size of the input space $|I|$ under exploration and 
(ii) the maximum number of executed instructions $N$ along the path (assuming no infinite loops, since such paths are non-exploitable and cannot be fuzzed). 
In the worst case, a monolithic, or traditional end-to-end fuzzer must explore all $|I|$ possible inputs and execute up to $N$ instructions for each, yielding a complexity of $O(|I| \cdot N)$ for validating the exploitability of a candidate path $\pi$.

\paragraphtitle{Complexity of Segmented Fuzzing}
If a path $\pi$ is decomposed into semantically-preserving segments $\{S_1, \ldots, S_k\}$, then a segmented fuzzing campaign explores each segment independently. 
Under this assumption, the total exploration cost becomes the \textit{sum} of the state spaces of each of its segments $O(|I_i| \cdot N_i)$.
%

Electron's multi-process architecture with Progression APIs at process boundaries creates natural opportunities for each of the three segmentation patterns below.
To illustrate these benefits, and without loss of generality, we focus on the simplest case: 
decomposing a path $\pi$ into two segments, $S_a$ and $S_b$. 
In the remainder of this subsection, we identify the principles under which such segmentation yields substantial reductions in explorable state space enabling efficient exploit validation, also summarized in Table~\ref{tab:complexity}.

\paragraphtitle{Dataflow-Independent Segments}
The first pattern arises when two segments receive inputs from \textit{independent} external sources. For example, in the Paperlib case, Source U (the custom URI) and Source S (the CSRF-fetched server response) are independent. The content of each is controlled separately by the attacker. A monolithic fuzzer must explore their full cross-product $|I_a| \times |I_b|$; segmentation reduces this to $O(|I_a| \cdot N_a) + O(|I_b| \cdot N_b)$, collapsing the input space from multiplicative to additive size.


\paragraphtitle{Partially Dependent Segments}
The second pattern occurs when $S_b$ receives a mixture of inputs: some derived from $S_a$'s output, and some from an independent source.
Electron's IPC exemplifies this: the channel field is determined by $S_a$ (fixing routing), while the content field can be an arbitrary attacker-controlled value.
Once valid channel bindings are known, the content space $I_b$ can be fuzzed independently.
Let $I_a' \subseteq I_a$ be the subset of $S_a$'s inputs that produce valid bindings. The computation of speedup can be found in Table~\ref{tab:complexity}.



\paragraphtitle{Fully Dependent Segments}
The final pattern arises when $S_b$'s input is \textit{entirely} determined by $S_a$'s output, and $S_a$ performs an \textit{invertible} transformation, which is common in Electron apps where attacker-supplied data is first parsed via \texttt{new URL}, \texttt{JSON.parse}, or \texttt{decodeURIComponent}.
Invertibility means that given any valid parsed output, a corresponding input can be reconstructed, enabling \sys to bypass $S_a$ entirely and fuzz directly over valid parsed values.
In the Paperlib example, Segment A parses the custom URI via \texttt{new URL} and \texttt{JSON.parse}. \sys skips fuzzing Segment A and instead uses an input reconstructor to generate well-formed URIs backward from valid parsed outputs. 

\subsection{Static Progression Analysis}
\label{subsec:static}

\sys's static progression analysis takes the Electron application source code as input and produces, for each discovered exploitable path $\pi$: its segmentation $\text{seg}(\pi) = \{S_1, \ldots, S_k\}$, per-segment fuzzing harnesses, and LLM-generated seeds.
To achieve this, \sys employs an LLM agent that performs a comprehensive taint-style analysis over the target repository, guided by a system prompt (Appendix~\ref{sec:app-agent}) that provides three inputs: (1) a threat model and vulnerability description, (2) a catalog of Progression APIs and terminal sinks, and (3) structured analysis tasks defining what the agent must discover and produce.

\paragraphtitle{Threat Model and Vulnerability Description}
The agent is given a precise definition of \shortvul as a source-to-sink exploit chain originating from external sources such as custom URI and propagating through multi-process data flows, IPC channels, and transformation layers. The key is that \shortvul might chain multiple intermediate vulnerabilities, such as CSRF, JavaScript code execution, before it finally reaches an OS command injection.

\paragraphtitle{Catalog of Progression APIs and Terminal Sinks}
The agent is given the catalog shown in Table~\ref{tab:sinks}, which enumerates known Progression APIs, which include IPC mechanisms, external fetch operations, navigation APIs, and conditional code execution APIs, along with terminal sinks, across Electron's processes and execution contexts.
On the source side, the catalog includes the three payload source types (U, S, F) with their corresponding code patterns (e.g., \texttt{app.on('open-url')}, \texttt{fetch(T)}, \texttt{fs.readFile(T)}) that serve as taint entry points.
This catalog determines which APIs the agent flags as boundaries or exploitation targets during path analysis.

\paragraphtitle{Structured Analysis Tasks}
The agent is instructed to search the entire codebase, locate all potential sources and sinks, trace dataflows across process boundaries, and group them into segments along the segmentation boundaries.
For each discovered path, the agent must produce:
(1) an \texttt{analysis.md} summarizing identified sources, sinks, and segment boundaries such as IPC communications;
(2) a \texttt{dataflow-segments.json} enumerating each segment, its code location, process type, and linkage to neighboring segments;
(3) per-segment fuzzing harnesses; and
(4) LLM-generated seeds for each segment's initial corpus. We will detail the harness and seed generation in~\secref{subsec:fuzzing}.

\subsection{Segmented Fuzzing}
\label{subsec:fuzzing}

Given the segments, harnesses, and initial seeds produced by static progression analysis, segmented fuzzing takes each segment $S_i$ as input and produces a set of concrete inputs that either reach a terminal sink or an intermediate sink, that is, a Progression API, within that segment.
\sys fuzzes segments sequentially using three components: a fuzzing loop that drives execution, harnesses that serve as segment entries, and oracles that define success conditions.

\paragraphtitle{Fuzzing Loop}
Algorithm~\ref{alg:staged-fuzzing} presents the segmented fuzzing workflow.
For each segment $S_i$, \sys first checks whether it contains only invertible operations (lines 5--7); if so, the segment is skipped during fuzzing and reconstructed later using the input reconstructor.
For non-invertible segments, \sys instruments the program with the segment-specific harness and oracle, incorporating inputs from all previous segments (line 8), then executes fuzzing with the provided seeds (line 9). The \texttt{fuzz} function returns $\textit{inputs}_i$ containing the inputs that successfully trigger the target segment oracle.
If no valid inputs are found (lines 10--12), \sys terminates chain exploration, as subsequent segments cannot be reached.
Otherwise, if the previous segment $S_{i-1}$ was invertible (lines 13--16), \sys reconstructs its input by inverting from the current segment's inputs and appends it to the chain.
\sys then appends the current segment and its inputs to the chain (line 17).
Finally, \sys returns the complete chain (line 19), which contains the inputs needed to trigger each segment in the detected exploit path.

\begin{algorithm}[t]
\caption{Scaffold of segmented fuzzing workflow}
\label{alg:staged-fuzzing}
\begin{algorithmic}[1]
\STATE \textbf{Input:} Program $P$, Segments $S_i$, Harnesses $\textit{harness}_{1:k}$, Oracles $\textit{oracle}_{1:k}$, Seeds $\textit{seeds}_{1:k}$, Input Reconstructors $\textit{invert}_{1:k}$
\STATE \textbf{Output:} Valid segment chains with inputs
\STATE $\textit{chain} \gets []$
\FOR{$i = 1$ to $k$}
    \IF{$S_i$ is invertible}
        \STATE \textbf{continue}
    \ENDIF
    \STATE $P^{(i)}  \gets \text{instrument}(P, \textit{harness}_i, \textit{oracle}_i, \textit{inputs}_{1:i-1})$
    \STATE $\textit{inputs}_i \gets \text{fuzz}(P^{(i)}, \textit{seeds}_i)$
    \IF{$\textit{inputs}_i = \emptyset$}
        \STATE \textbf{break}
    \ENDIF
    \IF{$S_{i-1}$ is invertible}
        \STATE $\textit{inputs}_{i-1} \gets \textit{invert}_{i-1}(\textit{inputs}_i)$
        \STATE $\textit{chain}.\text{append}((S_{i-1}, \textit{inputs}_{i-1}))$
    \ENDIF
    \STATE $\textit{chain}.\text{append}((S_i, \textit{inputs}_i))$
\ENDFOR
\RETURN $\textit{chain}$
\end{algorithmic}
\end{algorithm}

\paragraphtitle{Corpus}
Each segment's initial corpus comes from (a) the upstream segment's Progression API outputs. Every observed output contributes its concrete argument and a taint annotation describing the attacker-controlled substring, and (b) LLM-generated seeds from static analysis, producing one representative input per parse branch.
The first segment's corpus is synthesized from the attacker model.

\paragraphtitle{Halting}
A segment's fuzzing campaign halts when any of the following conditions are met:
(i) a terminal sink is reached, triggering the oracle and saving a proof-of-concept;
(ii) coverage of the segment's recognized branches saturates and no new Progression API output is observed in $k$ iterations; or
(iii) a global time budget expires.
A global halt condition additionally fires when a terminal sink is reached in any segment reachable from $S_1$ via a composable sequence of Progression API outputs, which constitutes a complete end-to-end proof-of-concept.

\paragraphtitle{Harness Synthesis}
%
%
For each segment $S_i$, \sys synthesizes a lightweight, in-place testing harness that directly triggers $\text{src}(S_i)$ with a fuzzer-supplied byte buffer. 
The harness is generated automatically from the segment specification produced during static analysis. 
It reconstructs only the minimal program state needed for the segment, replaying the source (such as an event handler, IPC entry point, or function call) without executing prior segments. 

Each harness is materialized as a small source-code patch that registers a function under a \texttt{global} namespace, enabling the fuzzing engine to invoke it directly.
The synthesized harness maps input bytes into the structured data expected by that segment (e.g., a URI string, a parsed object, an IPC message payload) while hard-coding irrelevant contextual values. 
For segments in renderer processes, \sys modifies the window initialization so that Node.js APIs are available and the fuzzing runtime can be loaded. 
Applied patches leave the application’s normal behavior intact while providing precise, segment-level fuzzing entry points.

\paragraphtitle{Oracle Design}
\sys detects whether fuzz inputs successfully reach segment sinks $\text{snk}(S_i)$ using two oracle strategies.
\textit{Syntax-based oracles} monitor parse-time or evaluation-time syntax errors emitted by APIs that validate their inputs.
Such errors are strong indications that the fuzzer-generated input reached the API, broke out of its original context, and was parsed as code, following prior works~\cite{trickel2023toss, guler2024atropos}.
\textit{Canary-based oracles} embed a unique marker into fuzz inputs and check for its presence in API arguments or side effects, including network navigation targets, file paths, and IPC messages. Detection of the canary confirms that input propagation reached the intended sink.
An example of such a canary is depicted on the bottom-right corner of Figure~\ref{fig:motivating}.

\subsection{End-to-end Validation}
\label{subsec:validation}

Finally, given the validated segments and their respective inputs, \sys performs end-to-end validation through two steps: payload composition and end-to-end testing.

\paragraphtitle{Payload Composition}
\sys first composes the final exploit payload through backward propagation from the last segment to the first. 
For each segment $S_i$ (traversing from $k$ to $1$), \sys maps the fuzzing harness inputs to the original program variables of $S_i$, then embeds this reconstructed data into segment $S_{i-1}$'s canary placeholders to propagate the payload backward. Note that all segments except the last contain canary values in their inputs, as only the final segment's input can trigger the syntax-based oracles.
For example, in Figure~\ref{fig:motivating}, \texttt{harness\_C}'s inputs  \texttt{buf} are mapped back to \texttt{msg.data} as the array \texttt{["import-from", "https://\$canary"]} based on line 2 of the utility process code. Then, for Segment B, \sys replaces the canary in its fuzzing input \texttt{args} with this reconstructed array value. This process continues backward: Segment B's \texttt{args} value is embedded into Segment A's URL parameter. 

\paragraphtitle{End-to-end Testing}
\sys then executes the composed payload against the complete application to verify the exploit path works end-to-end. 
When the exploit chain involves multiple sources, \sys instruments the application to intercept and supply secondary payloads directly. Specifically, \sys inserts condition blocks that check if a requested resource (domain or file path) is valid and matches the source specified in the primary payload. If so, it returns the secondary source's payload instead of performing the actual fetch or file read. 
%





\begin{table*}[!t]
\centering
\scriptsize
\caption{[RQ1] A selective list of zero-day \shortvuls detected by \sys. 
The column ``Payload Sources" indicates the attack vectors. 
The ``Consequence" column shows exploitation impact. 
The ``Status" column indicates disclosure status: Fixed (patch released), Acknowledged (maintainer confirmed but not yet fixed), or Reported (under review). 
}
\label{table:rq1-zero-day}
\setlength{\tabcolsep}{2pt}
\renewcommand{\arraystretch}{1.1}
\begin{tabular}{
  >{\arraybackslash}p{1.8cm}      
  >{\arraybackslash}p{1.8cm}  
  >{\centering\arraybackslash}p{0.7cm}  
  >{\centering\arraybackslash}p{1.6cm}  
  >{\centering\arraybackslash}p{1.7cm}  
  >{\centering\arraybackslash}p{1.8cm}  
  >{\centering\arraybackslash}p{3.6cm}  
  >{\centering\arraybackslash}p{1.4cm}  
  >{\centering\arraybackslash}p{2.2cm}  
}
\toprule
    \textbf{Application} &
    \textbf{Category} &
    \textbf{Stars} &
    \textbf{Version} &
    {\shortstack{\textbf{Payload (U/F/S)}}}&
    \textbf{Safe Config.} &
    \textbf{Exploit Chain} &
    \textbf{Status} &
    \textbf{CVE/Advisory} \\
\midrule
AFFiNE           & Note    & 59.7k & v0.25.1   & U   & \yes     & LFI $\rightarrow$ RCE & Fixed        & CVE-2026-XXXX  \\
Motrix           & Download Mgr.     & 49.8k & v1.8.19         & U      & \yes 
& AFW & Reported     & --             \\
Hyper            & Terminal          & 44.5k & v4.0.0-canary.5 & U      & \yes 
&  RCE & Acknowledged & --             \\
Cherry Studio    & AI Client         & 35.4k & v1.4.11         & U      & \yes  
& JSE $\rightarrow$ RCE & Fixed        & CVE-2025-XXXX  \\
Mailspring       & Email Client      & 16.8k & v1.16.0         & U      & \yes 
& RCE & Reported     & --             \\
DevHub           & Dev Tool          & 10.0k   & v4.0.3          & U      & \yes 
& JSE $\rightarrow$ RCE & Reported     & --             \\
MusicFreeDesktop & Music Player      & 7.0k    & v0.0.8          & U+S & \yes 
& RCE & Acknowledged & CVE-2025-XXXX  \\
Pinokio          & AI App Launcher   & 5.7k  & v3.9.0          & U+F    & \no 
& CSRF $\rightarrow$ LFI $\rightarrow$ RCE & Fixed        & CVE-2025-XXXX  \\
Deepchat         & AI Client         & 4.9k  & v0.3.0          & U      & \yes 
& DOM XSS $\rightarrow$ RCE & Fixed        & CVE-2025-XXXX  \\
LBRY Desktop     & Media Streaming   & 3.5k  & v0.53.9         & U+F    & \no
& LFI $\rightarrow$ RCE & Reported     & CVE-2025-XXXX  \\
Eidos            & AI Data Mgr. & 3.0k    & v0.21.0         & U      & \yes 
& JSE $\rightarrow$ SBE $\rightarrow$ RCE & Fixed        & CVE-2025-XXXX  \\
Thorium Reader   & E-book Reader     & 2.4k  & v3.2.2          & U+S & \yes 
& CSRF $\rightarrow$ LFI $\rightarrow$ RCE & Fixed        & GHSA-XXX-XXX \\
Paperlib         & Reference Mgr.    & 2.0k    & v3.1.10         & U+S & \no 
& CSRF $\rightarrow$  LFI $\rightarrow$ DOM XSS $\rightarrow$ RCE & Fixed        & CVE-2025-XXXX  \\
TidGi-Desktop    & Note    & 1.9k  & v0.12.4         & U      & \yes 
& JSE $\rightarrow$ RCE & Fixed        & GHSA-XXX-XXX \\
Muffon           & Music Streaming   & 1.9k  & v2.2.0          & U+S & \yes
& CSRF $\rightarrow$ DOM XSS $\rightarrow$ RCE & Fixed        & CVE-2025-XXXX  \\
Dive             & Docker Tool       & 1.6k  & v0.9.3          & U      & \yes
& JSE $\rightarrow$ RCE & Fixed        & CVE-2025-XXXX  \\
Vieb             & Web Browser       & 1.5k  & v12.3.0         & U      & \no
& LFI $\rightarrow$ RCE & Fixed        & GHSA-XXX-XXX \\
nanovault        & Crypto Wallet     & 186   & v1.2.1          & U      & \yes 
& JSE $\rightarrow$ RCE & Reported     & CVE-2025-XXXX  \\
\bottomrule
\end{tabular}
\begin{flushleft}
\textbf{Payload:} U=(Custom) URI, F=File System, S=Remote Server;
\textbf{Safe Config.}:
\yes: the exploitation works under the default safe configuration (\texttt{nodeIntegration=false}, \texttt{contextIsolation=true}, and \texttt{sandbox=true});
\no: otherwise;
\textbf{Exploit Chain}:
CSRF: Client-side Request Forgery;
DOM XSS: DOM Cross-site Scripting;
SBE: Sandbox Escape;
JSE: JavaScript Code Execution;
LFI: Local File Inclusion;
RCE: OS-level Remote Code Execution.
\end{flushleft}
\end{table*}

\section{Implementation}
\label{sec:implementation}

\sys implements specialized fuzzing workflows using segmented fuzzing, targeting multiple input sources, multiple processes, and diverse harnesses, and introduces engine-level canary detection to automatically identify when attacker-controlled inputs reach security-sensitive APIs.
Building upon jazzer.js~\cite{codeintelligence2023jazzerjs}, a JavaScript fuzzing framework based on libfuzzer~\cite{libfuzzer}, the implementation consists of 1,130 lines of code changes modifying the Electron engine, 948 lines extending jazzer.js with additional instrumentation, and 1,012 lines of agentic workflows.
To facilitate future research, we have open-sourced \sys at \repourl.

\paragraphtitle{Static Analysis} 
\sys implements agentic static analysis using Claude Opus 4.7 as the underlying LLM, while Claude Code serves as the agentic framework. 
The agent operates with a standard toolchain including file reading, file writing, directory traversal, and code search functionalities.

\paragraphtitle{Fuzzing}
\sys instruments targets at two levels. At the framework level, it modifies Electron's source and the Node.js standard library to hook APIs serving as segment oracles (sinks in Table~\ref{tab:sinks}) and external input APIs. At the application level, it injects harness code and JavaScript-specific coverage feedback. For main and utility processes, \sys hooks Node.js module loading for runtime instrumentation; for renderer processes, it instruments code at compile time via build plugins (e.g., \texttt{vite.config.ts}).


\paragraphtitle{Validation}
\sys uses an LLM agent with the standard toolchain extended with a tool to launch applications with custom URI payloads for end-to-end payload validation.

\section{Evaluation}
\label{sec:evaluation}
We systematically evaluate \sys through the following five research questions (RQs):
\begin{icompacter}
\item \textbf{RQ1 [Zero-days]}: How many zero-day MPVs does \sys discover in real-world Electron applications, and what is the impact of these vulnerabilities?
%
\item \textbf{RQ2 [Accuracy]}: What are the false positive and false negative rates of \sys ?
\item \textbf{RQ3 [Performance]}: How does \sys perform across its two phases? In Phase I, how many candidate paths and segments are produced per application, and how long and how many tokens does static analysis take? In Phase II, how long does fuzzing each segment take in terms of time-to-exposure (TTE)?
%
\item \textbf{RQ4 [LLM Ablation]}: What is the contribution of LLM-generated harnesses and seeds to \sys's detection capability? Specifically, how does \sys perform when LLM-generated harnesses and seeds are replaced with templates or random ones?
\item \textbf{RQ5 [Segmentation Ablation]}: How does \sys's segmented fuzzing compare to monolithic end-to-end fuzzing?
\end{icompacter}

\subsection{Experimental Setup}

\paragraphtitle{Dataset}
We collected open-source Electron applications from GitHub by identifying repositories with \texttt{electron} listed as a dependency in \texttt{package.json} and over 100 stars, spanning projects from Jan. 2010 to April 2026.
To focus on applications with external sources such as custom URI handling, we filtered for repositories containing \texttt{"open-url"} or \texttt{"second-instance"} in their codebase.
This yielded \datasetsize Electron applications. We will release the dataset upon acceptance.

\paragraphtitle{Environment}
All experiments were conducted on a Mac workstation with an Apple M4 Pro processor and 24 GB of memory, running macOS 15.7.2.
\sys uses \texttt{claude-opus-4-7} for agentic static analysis.

\subsection{RQ1: Zero-day Vulnerabilities}

In this subsection, we answer the research question regarding the zero-day vulnerabilities detected by \sys.
We define a detected vulnerability as a zero-day \vul vulnerability if there is no prior public disclosure and it is confirmed by a human expert with a successful end-to-end exploit.
In total, \sys discovered \zeroday zero-day vulnerabilities: \zerodayrce leading to remote code execution and one enabling arbitrary file write.

Table~\ref{table:rq1-zero-day} presents a selective list of those zero-day vulnerabilities, including the affected applications' details, payload sources, vulnerability consequences, and disclosure status.
Many of the affected applications are highly popular, with up to 60k stars, demonstrating that \shortvul vulnerabilities affect widely-used software.
The vulnerable applications span diverse categories including note-taking software, AI desktop assistants, email clients, music players, browsers, and document readers. This diversity indicates that \shortvul vulnerabilities are not limited to specific application types but represent a systemic risk across the Electron ecosystem.
%
We responsibly disclosed all vulnerabilities through GitHub Security Advisories or direct email to maintainers. Of the \zeroday vulnerabilities, \cve have been assigned CVEs, and \ack have been acknowledged by maintainers, with \fix already fixed to date.
We were also rewarded a bug bounty by Vercel for the discovery of the \shortvul vulnerability in Hyper.

The discovered vulnerabilities exhibit diverse vulnerability code patterns across payload sources and sink contexts.
%
%
%
%
Among them, four vulnerabilities require Remote Server Response, where the URI triggers the application to fetch and process content from attacker-controlled servers, such as downloading plugins (MusicFreeDesktop), retrieving metadata (Muffon), downloading ebooks (Thorium Reader), or fetching references (Paperlib).
Two vulnerabilities require local file access, where the URI triggers applications to load a local file via dangerous APIs, such as requiring JavaScript files via \texttt{require} (Pinokio) or loading local HTML via \texttt{window.loadURL} (LBRY Desktop).
%
%
Table~\ref{table:rq1-zero-day} also indicates whether exploitation succeeds under the default safe configuration (\texttt{nodeIntegration=false}) in the Safe Config.\ column. Entries marked \yes exploit the application through message progression rather than relying on unsafe configurations, distinguishing our work from prior work focused on unsafe configuration-caused issues~\cite{ali2024rise}.

\subsection{RQ2: Accuracy}

\sys's detection pipeline operates in two phases.
Phase I (agentic static analysis) flagged \llmreport out of \datasetsize applications as containing potentially exploitable paths.
Phase II (segmented fuzzing) then fuzzes the segments in each flagged application, producing \phasetwo apps with proofs-of-concept that reach a terminal sink.
Of the remaining flagged applications, fuzzing either timed out within the four-hour budget or failed to reach a terminal sink.
Of the \phasetwo confirmed cases, \zeroday lead to either OS-level Remote Code Execution (RCE) or Arbitrary File Write (AFW).
We evaluate the accuracy of Phase II reports below.

\paragraphtitle{False Positives}
\sys considers a path exploitable when end-to-end validation succeeds with a working proof-of-concept that triggers the sink canary, indicating that the input achieves code execution in one of the main, utility, or renderer processes.
Under \sys's threat model, we only consider vulnerabilities that lead to RCE or AFW as true positives.
Since renderer process code execution may require additional exploitation steps that cannot be easily validated through oracles, \sys conservatively reports all code execution instances for manual evaluation.

Among the \phasetwo apps that \sys fuzzes to reach a terminal sink, manual expert evaluation determined that three of them cannot be escalated to RCE or AFW, yielding a false positive rate of 11.5\% (3/\phasetwo). 
The three cases reach the terminal sink but can not compose into valid end-to-end exploits. Two reach \texttt{openExternal} as the terminal sinks, while the code filters the argument with \texttt{http} or \texttt{https} protocols. The other one reaches \texttt{new Function} as the terminal sink in a runtime context where \texttt{require} is not available. As a result, reaching the function still prevents it from executing arbitrary code.

\paragraphtitle{False Negatives}
Evaluating false negatives requires a ground truth dataset of known \shortvul, but no such dataset exists, and no prior tools exist for \shortvul detection.
To assess potential false negatives, we take three complementary approaches.
First, we conducted an extensive search across the NVD database, GitHub security advisories, security disclosure platforms, and technical blogs, yielding only one publicly documented real-world \shortvul case~\cite{spaceraccoon2020open} that is triggered by a fully external resource like a custom URI.
The vulnerability exploits a redirection chain: the attacker causes the Electron application to load a website vulnerable to redirection, then the redirected attacker-controlled page calls \texttt{require} directly (because the renderer has \texttt{nodeIntegration} enabled). 
We ran \sys on this case and confirmed it successfully detects the vulnerability, by reporting the vulnerable path, and reaches \texttt{loadURL} as a terminal sink during fuzzing.
Second, we randomly sampled 30 applications from the \llmreportfp flagged by Phase I that \sys did not confirm as exploitable during Phase II fuzzing, and had two experts manually inspect their code for RCE or AFW vulnerabilities.
None of the 30 were confirmed exploitable.
Third, we randomly sampled 30 applications from the \unflagged applications not flagged by Phase I and manually inspected them for exploitable \shortvul.
Together, these three assessments suggest that \sys achieves low false negative rates across both phases of its pipeline.

\subsection{RQ3: Performance}
\begin{table}[t]
\centering
\caption{
    [RQ3] Average time and token cost for static analysis and harness generation for vulnerable projects with 100k+ Lines of Code (LoC). All times are in minutes.
}
\label{tab:performance-agent}
\resizebox{\columnwidth}{!}{%
\begin{tabular}{lcccc}
\toprule
\textbf{App} & \textbf{LoC}  & \textbf{Taint Ana. Time} & \textbf{Harness Gen. Time} & \textbf{Tokens (k)} \\
\midrule
Hyper         & 162k  & 6.10 & 4.75 & 146 \\
Cherry Studio & 140k  & 5.55 & 3.77 & 150 \\
Muffon        & 129k  & 5.90 & 5.05 & 133 \\
Eidos         & 120k  & 6.90 & 4.30 & 156 \\
Paperlib      & 119k & 5.08 & 8.40 & 157 \\
%
Median (I)    & 88k & 6.03 & 8.07 & 141 \\
Median (ALL)  & 65k   & 2.42 & 7.72 & 126 \\
\bottomrule
\end{tabular}%
}
\footnotesize
\begin{flushleft}
I: \llmreport Phase I report vulnerable apps;
ALL: \datasetsize total apps.
\end{flushleft}
\end{table}

We evaluate \sys's performance across its two phases: agentic static analysis (Phase I) and segmented fuzzing (Phase II).

\begin{figure}
    \centering
    \footnotesize
    \definecolor{histblue}{RGB}{149,196,235}
    \definecolor{histred}{RGB}{252,210,225}
    \begin{subfigure}[t]{0.48\linewidth}
        \centering
        \begin{tikzpicture}
            \begin{axis}[
                ybar,
                width=1.15\linewidth,
                height=4.2cm,
                bar width=0.35cm,
                enlarge x limits=0.20,
                xlabel={\#Dataflows},
                ylabel={\#Apps (log)},
                y label style={at={(axis description cs:-0.15,.5)}},
                xtick style={draw=none},
                xtick={0,1,2,3,4,5,6},
                x tick label style={font=\footnotesize},
                ymode=log,
                ymin=0.5, ymax=1000,
                ytick={1,10,100,1000},
                log ticks with fixed point,
                ymajorgrids=true,
                grid style={dashed, gray!40},
            ]
                \addplot[
                    fill=histblue, 
                    draw=histblue!40!black,
                    nodes near coords={\pgfmathprintnumber[fixed,precision=0]{\pgfplotspointmeta}},
                    nodes near coords style={font=\tiny, anchor=south},
                    point meta=rawy,
                ] coordinates {
                    (0,473) (1,61) (2,32) (3,15) (4,2) (5,5) (6,1)
                };
            \end{axis}
        \end{tikzpicture}
        \vspace{-5px}
        \caption{Dataflows per application.}
        \label{fig:flows-per-app}
    \end{subfigure}
    \hfill
    \begin{subfigure}[t]{0.48\linewidth}
        \centering
        \begin{tikzpicture}
            \begin{axis}[
                ybar,
                width=1.15\linewidth,
                height=4.2cm,
                bar width=0.21cm,
                enlarge x limits=0.15,
                xlabel={\#Segments},
                ylabel={\phantom{\#Apps (log)}},
                y label style={at={(axis description cs:-0.15,.5)}},
                xtick style={draw=none},
                xtick={0,2,4,6,8,10,12},
                x tick label style={font=\footnotesize},
                ymode=log,
                ymin=0.5, ymax=1000,
                ytick={1,10,100,1000},
                log ticks with fixed point,
                ymajorgrids=true,
                grid style={dashed, gray!40},
            ]
                \addplot[
                    fill=histred, 
                    draw=histred!40!black,
                    nodes near coords={\pgfmathprintnumber[fixed,precision=0]{\pgfplotspointmeta}},
                    nodes near coords style={font=\tiny, anchor=south},
                    point meta=rawy,
                ] coordinates {
                    (0,473) (1,11) (2,36) (3,24) (4,18) (5,4)
                    (6,12) (7,1) (8,3) (9,3) (10,3) (12,1)
                };
            \end{axis}
        \end{tikzpicture}
        \vspace{-5px}
        \caption{Segments per application.}
        \label{fig:segments-per-app}
    \end{subfigure}
    \vspace{-5px}
    \caption{[RQ3] Distribution of detected dataflows and segments per application ($n=\datasetsize$).}
    \label{fig:dist-flows-segments}
\end{figure}
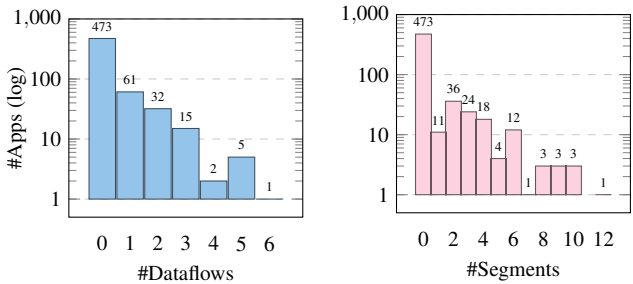

\paragraphtitle{Static Progression Analysis}
Phase I flagged \llmreport out of \datasetsize applications as potentially vulnerable, which produce 209 dataflows and 419 segments in total. Figure~\ref{fig:dist-flows-segments} shows the distribution of reported dataflows and segments per application across the full dataset. 
Among the \llmreport flagged applications, the majority (80.2\%) have one or two candidate dataflows. Each of the flagged applications has, on average, 1.80 candidate dataflows and 3.61 segments.
%

Table~\ref{tab:performance-agent} presents detailed metrics for selected vulnerable applications with over 100k LoC, including the number of segments, analysis time, harness generation time, and token consumption. Note that harness generation applies only to Phase I flagged applications. 
Among the \llmreport flagged applications, the median analysis time is around six minutes with a median token consumption of 141k tokens. 
Across all \datasetsize applications, the median analysis time is 2.42 minutes (this is because many do not contain vulnerable paths and the analysis stops early) with a median token consumption of 126k tokens. Even for the largest projects with over 100k LoC, the agentic static analysis is completed within sixteen minutes, showing its scalability.

\paragraphtitle{Segmented Fuzzing}
Table~\ref{tab:TTE-fuzz} shows the number of fuzzing segments, process types, their sources and sinks, and the time-to-exposure (TTE) for each segment across vulnerable applications. TTE measures the elapsed time from the start of fuzzing until a generated input triggers an oracle, that is, either reaching an intermediate sink or a terminal sink.
The median TTE across all vulnerabilities is {28} minutes, with {82\%} of vulnerabilities discovered within the 30 minutes of fuzzing. 
Single-segment vulnerabilities are resolved quickly. The shortest TTE of 13 seconds for Cherry Studio indicates cases where the vulnerability is directly reachable with minimal seed mutation. 
Multi-segment vulnerabilities take longer. Paperlib's four-segment chain requires 2h38m total, with one segment (M2: \texttt{rpc.postMessage}) accounting for 2h20m due to complex parsing constraints. 
We will discuss comparing the TTE to baseline fuzzing later.
%
%
\begin{table}[bhtp]
  \centering
  \scriptsize
  \caption{[RQ3] Per-segment and cumulative Time-to-Exposure (TTE) for multi-segment vulnerable paths under segmented fuzzing. 
  }
  \label{tab:TTE-fuzz}
  \resizebox{\columnwidth}{!}{%
  \begin{tabular}{p{1cm}clcc}
    \toprule
    \multirow{3}{*}{\textbf{Application}} 
    & \multirow{3}{*}{\textbf{Seg}} 
    & \multirow{3}{*}{\textbf{Inter./Terminal Sink}} 
    & \multicolumn{2}{c}{\textbf{\sys}} 
    \\
    \cmidrule(lr){4-5} 
    & & & \textbf{Seg TTE} & \textbf{Total TTE} \\
    \midrule
    \multirow{2}{*}{Deepchat}  & M1 & \texttt{sendToRenderer} & 8s  & \multirow{2}{*}{1h12min17s} \\
       & R1  & \texttt{v-html} & 1h12min9s\\
    \cmidrule(lr){1-5}
    \multirow{2}{*}{Dive} 
    & M1 &  \texttt{webContents.send} & 23s   & \multirow{2}{*}{9m55s}  \\
        & R1 &  \texttt{spawn}    & 9m32s & \\
    \cmidrule(lr){1-5}
    \multirow{2}{*}{Motrix} & M1 & \texttt{sendCommandToAll} & 21s &  \multirow{2}{*}{14m10s} \\
    & R1  & \texttt{aria2.addUri}  & 13m49s\\
    \cmidrule(lr){1-5}
    \multirow{3}{*}{muffon} 
        & M1 & \texttt{webContents.send()}    & 37s   & \multirow{3}{*}{3m59s}  \\
        & R1 & \texttt{openNewTab} & 2m27s &  \\
        & M2 & \texttt{v-html}     & 55s   &  \\
    \cmidrule(lr){1-5}
    \multirow{4}{*}{paperlib} 
        & M1 & \texttt{postMessage} & 33s     & \multirow{4}{*}{2h38m9s}  \\
        & R2 & \texttt{get}             & 17m20s  &  \\
        & U1 & \texttt{rpc.postMessage}              & 2h20m11s &  \\
        & R2 & \texttt{innerHTML}       & 5s      &  \\
    \cmidrule(lr){1-5}
    \multirow{3}{*}{pinokio} 
        & M1 & \texttt{loadNewWindow}            & 14s    & \multirow{3}{*}{13m40s} \\
        & R1 & \texttt{res.redirect(w)} & 43s    &  \\
        & M2 &  \texttt{require}     & 12m43s & \\
    \cmidrule(lr){1-5}
    \multirow{3}{*}{\shortstack[l]{TidGi\\Desktop}} & M1 & \texttt{executeJavaScript} & 1m17s & \multirow{3}{*}{41m19s} \\ 
    & R1 & \texttt{wikiOperationInServer} & 27m1s\\ 
    & M2 & \texttt{new Function()} & 13m1s\\ 
    \cmidrule(lr){1-5}
    \multirow{4}{*}{\shortstack[l]{Thorium\\Reader}} & M1 & \texttt{httpGet} & 32s & \multirow{4}{*}{3h25m16s} \\ 
    & M2 & \texttt{writeStream.pipe}& 2h49m15s \\ 
    & M3  & \texttt{webContents.send}  & 35m8s\\ 
    & R1  & \texttt{webview.setAttribute} & 21s \\ 
    \bottomrule
  \end{tabular}
  }
\end{table}

\paragraphtitle{Scalability Analysis}
Phase I and Phase II together take around {30 minutes} on average, with a maximum of {three} hours for applications with large codebases and complex attack chains.
This time cost, combined with full parallelizability across applications, makes \sys practical for large-scale \shortvul detection in Electron applications.

\subsection{RQ4: Contribution of LLM Components}
\sys uses LLMs for taint analysis and harness/seed generation. Since the static taint analysis is central to \sys's design, we ablate only the latter two against non-LLM baselines (template harnesses, random seeds), with results in Table~\ref{tab:rq4-llm-ablation}.

\paragraphtitle{Harnesses}
LLM-generated harnesses are critical for renderer process fuzzing, where applications implement heterogeneous IPC handlers that template-based harnesses cannot generalize across. Replacing them with templates reduces detection from {\zeroday/\zeroday} to {6/\zeroday} (random seeds) and {11/\zeroday} (LLM seeds), dropping {17} and {12} vulnerabilities, respectively, as many segments fail to initialize correctly and never reach their target sinks.

\paragraphtitle{Seeds}
Beyond harnesses, good seeds are essential for branch exploration. LLM-generated seeds leverage semantic understanding of the application's input format to produce structurally valid initial inputs. With LLM harnesses fixed, replacing LLM seeds with random seeds reduces detection to {18/\zeroday}; the five missed cases all involve complex parsing where random seeds fail to reach the first segment's oracle. LLM seeds also accelerate discovery, reducing average TTE from 30m22s to 17m9s.

\begin{table}[htb]
\centering
\caption{[RQ6] Contribution of LLM components in \sys. 
}
\label{tab:rq4-llm-ablation}
\setlength{\tabcolsep}{8pt}
\footnotesize
\begin{tabular}{@{}llccc@{}}
\toprule
\textbf{Harness} & \textbf{Seeds} & \textbf{Detected} & \textbf{Avg. TTE} & \textbf{Avg. Tokens} \\
\midrule
Template & Random  & {6}/\zeroday  & {40m12s}  & ---  \\
Template & LLM     & 11/\zeroday & {28m04s} & {47k}  \\
LLM      & Random  & {18}/\zeroday & {30m22s} & {109k} \\
\midrule
\textbf{LLM} & \textbf{LLM} & \textbf{\zeroday/\zeroday} & \textbf{17m9s} & \textbf{156k} \\
\bottomrule
\end{tabular}
\end{table}

\subsection{RQ5: Segmented vs. End-to-End Fuzzing}

Since no prior work exists for \shortvul~detection, we implement an end-to-end fuzzing baseline based on {jazzer.js}, to serve as a baseline. 
To ensure fair comparison, we instrument the baseline with identical oracles and initial seeds as \sys. Both approaches have a four-hour time limit for fuzzing per application.

\paragraphtitle{Zero-day Detection}
Among the \zeroday zero-day \shortvuls detected by \sys, the baseline only succeeded in 6, with the remaining 17 timing out. All baseline successes are single-segment vulnerabilities where the terminal sink is directly reachable from the first input boundary.
For other multi-segment vulnerabilities, the baseline fails to propagate inputs across process boundaries within the time budget. 
For example, it fails to reach the first \texttt{postMessage} sink in Paperlib within four hours, while \sys reaches the terminal sink in 2h38m via segmented fuzzing with an input reconstructor handling the invertible URI parsing.

\paragraphtitle{Edge Coverage}
Table~\ref{tab:RQ5-proton-vs-baseline} compares edge coverage achieved by \sys and the baseline after four hours across 53 sampled applications. \sys achieves substantially higher coverage on vulnerable multi-segment applications. For example, 98.2\% more edges on eidos. Specifically, the coverage improvement appears in multi-segment vulnerabilities, consistent with the complexity analysis in \secref{subsec:insight}: segmentation allows \sys to explore deeper program states across processes, within the same time budget.

\begin{table}[tbhp]
  \centering
  \scriptsize
  \caption{[RQ5] Edge coverage comparison of \sys against baseline, over a selective list of apps reported with candidate paths with multiple segments.}
  \label{tab:RQ5-proton-vs-baseline}
  \resizebox{\columnwidth}{!}{%
  \begin{tabular}{llll}
    \toprule
    \textbf{Application} &   \textbf{Baseline EdgeCov@4h}   & \textbf{\sys EdgeCov@4h} & \textbf{$\Delta$}   \\
    \midrule
    Deepchat & 1,646 & 1,750 & +6.3\% \\
    \cmidrule(lr){1-4}
    eidos &  685 & 1,358 &  +98.2\%  \\ 
    \cmidrule(lr){1-4}
    Mailspring  & 345 & 772 & +123.8\%\\
    \cmidrule(lr){1-4}
    muffon &  389 & 433 & +11.3\%\\
    \cmidrule(lr){1-4}
    MusicFreeDesktop & 299 & 555 & +85.6\% \\
    \cmidrule(lr){1-4}
    Motrix  & 327 & 672 & +105.5\% \\
    \cmidrule(lr){1-4}
    paperlib & 525 & 1,150 & +119.0\%\\ 
    \cmidrule(lr){1-4}
    pinokio & 633 & 799 & +26.2\%\\
    \cmidrule(lr){1-4}
    Thorium Reader & 1,148 & 1,576 & +37.3\%. \\
    \cmidrule(lr){1-4}
    TidGi-Desktop & 594 & 1,128 &  +89.9\% \\
    \cmidrule(lr){1-4}
    median(S) & 477 & 892 & +87.0\% \\
    \bottomrule
  \end{tabular}
  }
 \footnotesize
\begin{flushleft}
S: 53 sampled apps, including \zeroday vulnerable ones and an additional 30 from phase I reports.
\end{flushleft}
\end{table}



\section{Discussion}
\label{sec:discussion}

\paragraphtitle{Mitigation}
Secure Electron configuration and other mitigations have been extensively studied~\cite{ali2024rise, jin2023security, xiao2022understanding, yang2025coindef}. 
On the handling side, defenses include: prompting users before handling external URIs, sanitizing inputs before dangerous APIs, validating domains via allowlists, isolating untrusted content in webviews, and configuring renderers with safe configurations enabled. 
On the triggering side, browsers and other applications that can invoke custom URIs should display the full URI string in confirmation dialogs, to enable users to identify malicious URIs.

\paragraphtitle{Generalizability of Segmented Fuzzing}
Segmented fuzzing generalizes beyond Electron to any multi-process target. The segmentation opportunities, including IPC boundaries, external input entry points, and invertible transformations, are broadly applicable. Porting requires adapting the agentic analysis to identify target-specific boundary APIs and implementing framework-specific instrumentation for harnesses, oracles, and reconstructors.



\paragraphtitle{Limitations}
\sys's agentic static analysis may miss complex dataflows when they span non-standard libraries, template engines, or LLM SDKs that require modeling beyond core Node.js and Electron APIs. LLM reasoning over such code may be incorrect due to limited training on recent libraries, inaccessible native addon source, and context window constraints. These inherent limitations we leave to future work.


\section{Related Work}
\label{sec:related_work}
%

\paragraphtitle{Security of Electron Applications}
Prior work on Electron security has focused on two main attack vectors. First, several studies~\cite{doyensec2022electronegativity, ali2024rise} examined vulnerabilities arising from unsafe configurations. Second, other works~\cite{xiao2022understanding, jin2023security, yang2025coindef} investigated malicious in-app user inputs and studied the defense of such vulnerabilities. 
%
%
%
In contrast, the exploitation of \vul, which needs to chain multiple vulnerabilities through cross-process messaging, has largely been ignored with no academic investigations.



\paragraphtitle{Web Application Fuzzing}
Traditional web fuzzers employ grey-box coverage feedback~\cite{witcher,luo2024banditimproveoddsoffuzzer,kim2024AIMFuzz}, data flow tracking~\cite{wang2024GuidedFuzzing,park2025RGFuzz,kim2025Chimera}, predicate synthesis~\cite{zhu2025locusagenticpredicatesynthesis,wang2025predator,lee2024syzrisk}, and LLM-generated grammars~\cite{meng2024large,dong2025fuzztestingmeetLLM,zhuo2024liftfuzzer} to explore deep program states. However, these fuzzers assume a single-process model~\cite{liu2024kernelfuzzing,eom2024fuzzingjavascript,zhu2024crossfire} and cannot be easily adapted to Electron's multi-process architecture. Segmented fuzzing via program slicing~\cite{chen2022Sfuzz}, function-level harnesses~\cite{Murali_2024}, or protocol modeling~\cite{feng2021snipuzzblackboxfuzzingiot,jinsheng2022greyboxfuzzing,li2022snpsfuzzer} shares our isolation goal, but cannot trace IPC flows across Electron processes or synthesize harnesses.

\paragraphtitle{Agentic Program Analysis}
LLM-assisted static analysis systems such as IRIS~\cite{li2024iris}, LLMDFA~\cite{wang2024llmdfa}, MoCQ~\cite{li2025mocq}, and QLCoder~\cite{wang2025qlcoder} augment dataflow reasoning or synthesize vulnerability queries, but operate on single-process programs and static code graphs. Agentic frameworks, including GPT-4 Analyst~\cite{cheng2023gpt4gooddataanalyst,noever2023largelanguagemodelsfix}, LLM4Fuzz~\cite{shou2024llm4fuzzguidedfuzzingsmart}, FuzzGPT~\cite{deng2023largelanguagemodelsedgecase}, and Locus~\cite{zhu2025locus}, iteratively refine inputs or fuzzing predicates but assume monolithic execution contexts. 

\section{Conclusion}
\label{sec:conclusion}

In this paper, we present \sys, the segmented fuzzing framework for detecting \vul in Electron applications. \sys decomposes complex cross-process exploit paths into independently fuzzable segments, enabling efficient fuzzing.
We evaluated \sys against \datasetsize real-world Electron applications, discovering \zeroday zero-day vulnerabilities, 
including \zerodayrce exploitable that enable remote code execution, with \cve CVEs to date. 
Our findings demonstrate that the \vul is critical yet overlooked in Electron applications, and our open-source release of \sys helps developers identify these vulnerabilities.

\cleardoublepage

{\footnotesize \bibliographystyle{IEEEtran}
\bibliography{bib}}


\cleardoublepage
\appendices

\section{Open Science}
\label{sec:open}

We provide the following artifacts to support reproducibility and future research. All artifacts are available at \url{https://anonymous.4open.science/r/proton} and will remain accessible after the submission deadline.

\paragraphtitle{Proton Engine}
We release the full \sys implementation, including: (1) the patch for Electron framework modifications with engine-level instrumentation for canary detection, Progression API hooks, and IPC interception; (2) the extended \texttt{jazzer.js} fuzzing runtime with additional JavaScript-specific coverage feedback, segment harness registration, and oracle integration; and (3) the scripts that coordinate 
Phase I and Phase II across applications.

\paragraphtitle{LLM-based Agentic Workflow}
We release the complete agentic static analysis and harness synthesis workflow, including: (1) the system prompts for the taint analysis and segmentation agent and the harness synthesis agent (Appendix~\ref{sec:app-agent}); (2) the LLM-generated \texttt{analysis.md} and \texttt{dataflow-segments.json} outputs for all vulnerable applications; (3) the synthesized harness patch files (\texttt{<id>.\allowbreak harness.patch}) for each segment; and (4) the LLM-generated seeds for each segment's initial corpus.

\paragraphtitle{Dataset}
We release two datasets: (1) the complete list of all \datasetsize Electron applications collected from GitHub, including their repository URLs, versions, and star counts at the time of collection; and (2) the curated list of \zeroday vulnerable applications, including their confirmed exploit chains, proof-of-concept payloads, and CVE identifiers. For applications with pending disclosure, exploit details will be released following the 45-day remediation period.

\paragraphtitle{Vulnerability Disclosures}
All \zeroday zero-day vulnerabilities have been responsibly disclosed to the respective maintainers. Proof-of-concept exploits for patched vulnerabilities are included in the artifact. For vulnerabilities still under remediation, proof-of-concept details will be added to the artifact repository upon public disclosure.

\section{Ethical Considerations}
We ensure that our study adheres to standard ethical guidelines for security research. All analyzed code was collected from publicly available open-source GitHub repositories in compliance with the platforms' terms of service. No private or user-sensitive data was accessed.
All application testing was conducted offline on local machines, with no impact on online servers or production systems.
Regarding vulnerability disclosure, we responsibly reported all identified vulnerabilities to affected vendors and allowed a 45-day remediation period before public disclosure. As of this writing, we have received \ack acknowledgments and \fix fixes, and \cve CVE identifiers have been assigned.

\begin{table*}[h]
\centering
\caption{Categorization of IPC communication mechanisms across Electron's multi-process architecture. For each IPC type, we show: the communicating parties with directionality, and code patterns for both sender and receiver.}
\label{tab:ipc}
\scriptsize
\setlength{\tabcolsep}{3pt}
\begin{tabular}{@{}l@{\hspace{8pt}}l@{\hspace{10pt}}l@{\hspace{8pt}}l@{}}
\toprule
\textbf{IPC Channels} & 
\textbf{Parties} & 
\textbf{Sender Code Patterns} & 
\textbf{Receiver Code Patterns} \\
\midrule

\multirow{3}{*}{Pre-defined IPCs} 
& Renderer $\rightarrow$ Main & 
{\tt ipcRenderer.send(C, Msg)} & 
{\tt ipcMain.on(C, F);ipcMain.addListener(C, F)} \\
& Renderer $\leftrightarrow$ Main & 
{\tt R = ipcRenderer.invoke(C, Msg)} & 
{\tt ipcMain.handle(C, F)} \\
& Main $\rightarrow$ Renderer & 
{\tt webContents.send(C, Msg)} & 
{\tt ipcRenderer.on(C, F);ipcRenderer.addListener(C, F)} \\
\midrule

\multirow{2}{*}{Event Handlers} 
& Renderer $\rightarrow$ Main & 
{\tt window.open(URL)} & 
{\tt webContents.setWindowOpenHandler(F)} \\
& Renderer $\rightarrow$ Main & 
{\tt location.assign(URL)} & 
{\tt webContents.on("will-navigate", F)} \\
\midrule

Internal Protocols 
& Renderer $\leftrightarrow$ Main & 
{\tt window.location.href = P} & 
{\tt protocol.handle(P, F)} \\
\midrule

\multirow{2}{*}{Pre-defined Message Ports} 
& Main $\rightarrow$ Utility & 
{\tt child.postMessage(Msg)} & 
{\tt process.parentPort.on('message', F)} \\
& Utility $\rightarrow$ Main & 
{\tt proc.parentPort.postMessage(Msg)} & 
{\tt child.on('message', F)} \\
\midrule

Custom Message Ports 
& Any $\leftrightarrow$ Any$^{*}$ & 
{\tt port.postMessage(Msg)} & 
{\tt port.on(F); port.onmessage(F)} \\

\bottomrule
\end{tabular}
\begin{flushleft}
\texttt{C}=Channel name, \texttt{Msg}=Message data, \texttt{F}=Handler function, \texttt{P}=Protocol scheme, \texttt{R}=Return value. $\rightarrow$: One-way communication; $\leftrightarrow$: Bidirectional communication. $^{*}$\xspace: Includes Main-Utility (M-U), Main-Renderer (M-R), Utility-Renderer (U-R), Renderer-Renderer (R-R), and Utility-Utility (U-U) communications.
\end{flushleft}
\end{table*}

\begin{figure}[htb]
    \includegraphics[width=\linewidth]{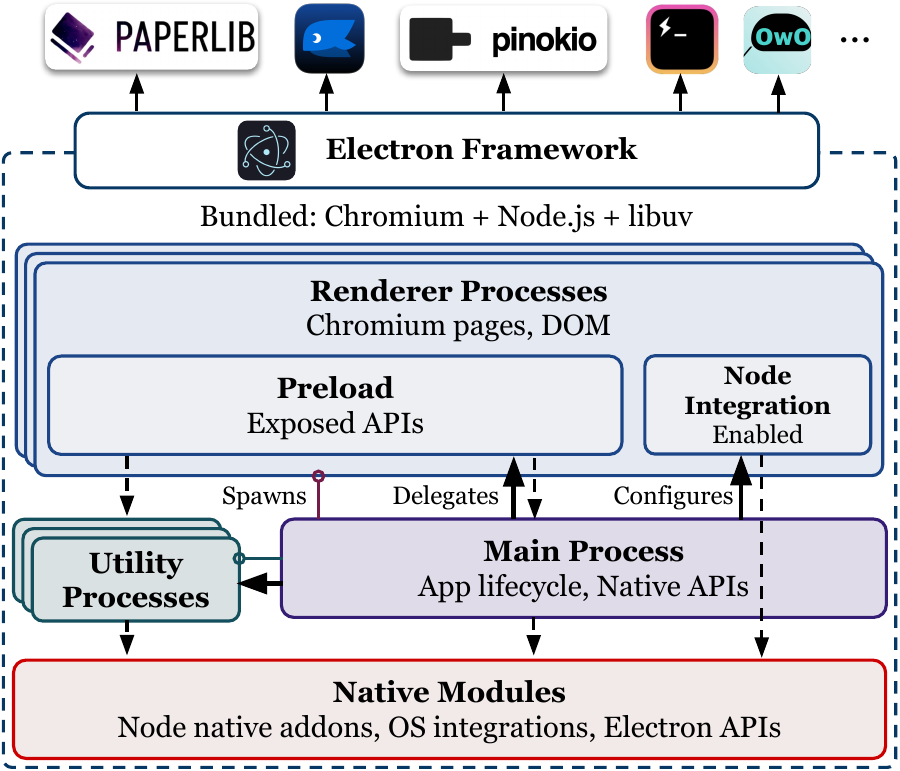}
    \caption{
        Electron architecture.
    }
    \label{fig:architecture}
\end{figure}

\section{Exploitation Techniques}
\label{sec:exploitation}

In this section, we explain the exploitation techniques used to escalate code execution in renderer processes to arbitrary command execution. 

\paragraphtitle{Unsafe Window Configuration}
Electron provides security configurations for renderer processes, including \texttt{contextIsolation} and \texttt{nodeIntegration}. Misconfigurations can enable code execution to escalate to command execution:

\begin{icompacter}
\item \textbf{Direct Access via nodeIntegration.} When node integration is enabled and context isolation is disabled, renderer processes have direct access to Node.js APIs. This is commonly seen in applications whose renderers need to use Node.js APIs.

\item \textbf{Prototype Pollution.} When context isolation is disabled but node integration remains disabled, the renderer and preload contexts share the same JavaScript environment, separated only by closure scope. Attackers can exploit this by overwriting built-in prototypes (e.g., \texttt{Function\allowbreak.prototype\allowbreak.apply}) to intercept function calls in the preload context and leak privileged objects. For example, in the Thorium Reader vulnerability, overwriting \texttt{Function.prototype.apply} allowed capturing the \texttt{ipcRenderer} object when preload scripts invoked \texttt{apply}. Once \texttt{ipcRenderer} is obtained, attackers can further hook \texttt{EventEmitter\allowbreak.prototype\allowbreak.emit} to leak the \texttt{process} object by triggering Node.js event system gadgets (e.g., by exceeding the max listener count to trigger warnings). With the leaked \texttt{process} object, attackers gain arbitrary command execution via \texttt{process.binding("spawn\_sync")}.
\end{icompacter}

\paragraphtitle{Preload Script APIs}
Applications often expose custom APIs from preload scripts through \texttt{contextBridge}. If preload scripts expose the \texttt{ipcRenderer} object or wrappers, attackers can invoke arbitrary IPC handlers in the main process. Vulnerable handlers that execute commands without validation can lead to direct command execution. Similarly, exposed APIs using Node.js functionality (e.g., file operations, shell execution) without sanitization can be abused for privilege escalation.

\paragraphtitle{Main Process Event Handlers}
Besides IPC handlers, Main processes can also register event handlers for renderer actions such as navigation and window creation. Navigation handlers like \texttt{\allowbreak setWindowOpenHandler} may call dangerous APIs such as \texttt{shell.\allowbreak openExternal} without validating destinations, enabling command execution through specially crafted URLs.

\paragraphtitle{V8 Vulnerability}
When \texttt{sandbox} is disabled and the application uses an outdated Electron version, attackers can exploit known V8 vulnerabilities to achieve arbitrary code execution, bypassing Electron security configurations.

\paragraphtitle{Script Pivot}
Finally, when direct escalation is not possible due to isolated contexts, attackers can pivot to more privileged renderer contexts. From a compromised iframe, attackers can use \texttt{postMessage} to achieve XSS in the parent context or force top-level navigation to load attacker-controlled content.

\section{Agent Design}
\label{sec:app-agent}

In this section, we detail the design of our static analysis agent as well as the harness synthesis agent.

\subsection{Taint Analysis and Segmentation Agent}
\label{subsec:app-analysis}

The taint analysis and segmentation agent performs two things simultaneously: one is to start from the open-url handler source to trace to an actual RCE sink, and the other is to produce dataflow segments.
The high-level prompt (with details removed to save space) is shown in Listing~\ref{lst:prompt-analysis}.
Note that the categorization of IPC communication machanisms across Electron's multi-process architecture (taxonomized in Table~\ref{tab:ipc}) is included in the prompt for the LLM agent to find precise segmentation boundaries.
One sample output pair of segment source and sink is depicted in Listing~\ref{lst:analysis-output}.
Empirically, typical agent trajectory starts from grepping the pattern of \texttt{open-url} to find the sources, and then iteratively trace through the dataflow to read more and more files.
At some point, we observe agent using grepping tool to find sink patterns by writing regexes that generalize our specified taxonomy, which further leads the agent to find effective exploitable paths.

\begin{figure*}
    \begin{minipage}{0.49\linewidth}
\begin{lstlisting}[style=mypromptstyle,caption={LLM prompt for static analysis and segmentation.},label={lst:prompt-analysis}]
You are a security analyst auditing Electron 
applications for unsafe open-url handling 
vulnerabilities. We have identified a class of 
vulnerability which we named Message Progression Vulnerability (MPV).
Assuming the Electron application is installed inside 
of a user's system, when the user click on a...

Due to the complex nature of Electron applications, the 
dataflow may not be linear--that is, it could be 
considered as a multi-staged dataflow with multiple 
segments. For instance, there could be a dataflow that 
starts within Main process, and through an IPC call, go
into the renderer process...

Your job is to systematically analyze the repository to 
find potential source and sink, and finding whether 
they could be connected via multiple segments in 
different processes...

At the end, please come up with two artifacts:
1. an `analysis.md` which contains the summary of your 
   analysis of the code base. List out the discovered 
   potential sources and sinks. Note that the goal is 
   not to find just a single dataflow; but multiple 
   dataflows with multiple potential sub-dataflow 
   segments.
2. an `dataflow-segments.json` which is a json list 
   that looks like the following
``` json
[ {
    "id": <a unique identifier>,
    "type": "source | sink",
    "process": "main | renderer | utility",
    "file_name": <the relative file name>,
    "line_num": <the line number>,
    "comment": <some comment about the node>,
    
    // if we have a source
    "source_type": "open-url-handler | uncontrolled-
        file-read | response-received | ...",
    "source": <exact code pattern of the source>,
    "connected_sink": <the id of the connected sink>,
    "introducing_new_taint": true | false,
    
    // if we have a sink
    "sink_type": "require | html-inject | js-inject | 
        ipc-call | xss | navigation | rce",
    "sink": <exact code pattern of the sink>,
    "connected_source": <the id of the source>
  }, ...]
```

# Additional Information
**1. Code Execution (XSS/RCE)**
- `executeJavaScript()`, `eval()`, `new Function()`
- `innerHTML`, `outerHTML`, `document.write()`, 
  `v-html`. Example payload: 
    `myapp://xss?code=<script>alert(1)</script>`
...
# Notes
- Please be comprehensive and search all javascript 
  related files including .js, .ts (javascript or
  typescript) as well as the hybrid files related to 
  front-end frameworks such as .vue or .jsx or
  .tsx for Vue and React applications...
- External taint sources could come from remote 
  requests (fetch(...)) or local file read (read(...))
  especially if it happens to be that the arguments or 
  urls to these functions are not fully controlled...
- When you see that there is a window.loadURL or 
  redirection with tainted URL or payload, you should
  find all the potentially connected process handlers 
  to start subsequent segments.

# Analysis Task
Your are now working on the `{app}` repository.
- source code directory: `./repos/{app}/`
- output directory: `./analysis-outputs/{app}/`
As instructed, please perform thorough security 
analysis on the `{app}` repository.
\end{lstlisting}
    \end{minipage}
    \hfill
    \begin{minipage}{0.49\linewidth}
\begin{lstlisting}[style=mypromptstyle,caption={LLM prompt for synthesizing fuzzing harness.},label={lst:prompt-harness}]
# Harness Generation
You are a security analyst auditing Electron 
applications for unsafe open-url handling vulns.
We have identified a class of vulnerability or attack 
which we named Message Progression Vulnerability (MPV). Assuming the Electron 
application is installed inside of a user's system, 
when the user click on a potentially compromised link 
in a browser, that URI may have a prefix of `my-app://` 
which is handled by the Electron app...

Due to the complex nature of Electron applications, the 
dataflow may not be linear--that is, it could be 
considered as a multi-staged dataflow with multiple 
segments. For instance, there could be a dataflow that 
starts within Main process, and through an IPC call, go
into the renderer process...

## Task Description
Your task would be that for each detected source come 
up with a testing harness. A testing harness at its 
core, would be a function that takes in a buffer or an 
uint8array. It should contain the code necessary to 
convert the buffer into structured input to continue 
the execution of the current segment, ignoring all the 
prior segments...

However, to make the testing harness actually runnable, 
we need to properly expose the testing harness. The 
actual synthesized harness should have the 
form of a diff patch, which specifies in which file and 
in which line we would like to remove lines or add new 
lines. Please name the file `<source_id>.harness.patch` 
for each source stated in the dataflow-segments.json.

## Harness Example
Here we give a concrete example of a harness that can 
be synthesized for our purpose:
```ts
export function urlFuzzInitialize() {
  (global as any).__proton__.harness = 
  function (buf: Buffer | Uint8Array): void {
    const constructedUrl = harness(buf);
    try {
      app.emit('open-url', makeOpenUrlEvent(), 
        constructedUrl);
    } catch (e) {
      // ...
    } } }
```
...
## Harness as Patches
Generate patches that look like the following:
```
--- a/main.ts
+++ b/main.ts
@@ -1,5 +1,6 @@
 function initialize() {
-    printf("Hello, world!\n");
+    urlFuzzInitialize();
 }
```

This file should be directly applicable by running git 
patch. It is okay to have multiple files being 
modified. The criteria should be that, once the 
application is loaded (all the modules being ready), 
there should be our fuzzing harness registered in
`global.__proton__`.

# Harness Generation Task
Your are now working on the `{app}` repository.
- source code directory: `./repos/{app}/`
- analysis directory: `./analysis-outputs/{app}/`
- output directory `./harness-outputs/{app}/`
Everytime you encounter a new source you should create 
`<source-id>.harness.patch`. The patch should be 
comprehensive such that an automated script can launch 
the app and start a fuzzing campaign using `jazzer.js`.
As instructed, please perform thorough security 
analysis on the `{app}` repository.
\end{lstlisting}
    \end{minipage}
\end{figure*}

\subsection{Harness Synthesis Agent}
\label{subsec:app-harness}

Given the segmented dataflow produced in \secref{subsec:static}, the next challenge is to instantiate each segment as a fuzzable entry point. For a given segment, fuzzing should begin directly at its sub-source so that fuzzing can focus on the semantics of the segment while avoiding redundant work and long execution chains.
To achieve this, \sys uses an LLM-driven agent to synthesize in-place fuzzing harnesses, each packaged as a patch file automatically inserted into the Electron application source code. 
The process is guided by a second structured prompt that provides the agent with:
\begin{icompacter}
    \item \textbf{Segmentation result}: \texttt{analysis.md} and \texttt{dataflow-segments.\allowbreak json} from the taint analysis, including all discovered sources, their process locations, and the sinks and subsequent sources they connect to.
    
    \item \textbf{Detailed task specification}: defining what a harness must do:
    take a \texttt{Uint8Array} from the fuzzer,
    convert it into the structured input expected at that segment's source, replay the segment’s entry point (e.g., invoking \texttt{app.emit("open-url", ...)}, calling an IPC handler, or directly invoking a function),
    and expose the harness to the fuzzing engine via \texttt{global.\_\_proton\_\_.harness}.

    \item \textbf{Constraints and requirements}:
    including where harnesses may be inserted, how to preserve functional behavior, how to handle renderer-process sources (by enabling \texttt{nodeIntegration}), and how to structure each harness patch so that it is directly applicable via git apply.
\end{icompacter}
The full prompt is shown in Listing~\ref{lst:prompt-harness}.

\paragraphtitle{Harness Generation Workflow}
For every source node in the segmentation output, \sys instructs the agent to generate a dedicated patch file named \texttt{<source-id>\allowbreak.harness\allowbreak.patch}. The patch contains all code edits needed to:
\begin{icompacter}
    \item \textbf{inject a new initializer} (e.g., \texttt{urlFuzzInitialize()} or a source-specific variant) early in the application startup path;
    \item \textbf{define the harness function} that maps bytes to the source’s expected input shape. The agent reuses surrounding program logic whenever possible, and may hard-code unrelated values to preserve determinism;
    \item \textbf{bind the harness} into the global namespace under \texttt{global.\allowbreak \_\_proton\_\_.harness}, allowing an external driver (\texttt{Jazzer.js}) to locate and execute it;
    \item \textbf{ensure the correct runtime environment}, which may include editing BrowserWindow creation to enable Node.js APIs or injecting \texttt{Jazzer.js} bootstrap code into renderer entry files.
\end{icompacter}

\paragraphtitle{Example Output}
Harness patches generated by \sys resemble conventional source patches (Listing~\ref{lst:harness-patch}).
These patches are self-contained: applying them to the repository yields an instrumented application that, when launched, automatically registers all synthesized harnesses. The fuzzing engine can then start a campaign by calling \texttt{global.\_\_proton\_\_.harness(buf)}, which replays the sub-source with a fuzzed payload.

\begin{figure}[!hbt]
\begin{lstlisting}[style=mypromptstyle,caption={Sample JSON output from agentic static analysis.},label={lst:analysis-output}]
[{
  "id": "src-1-main-openurl-handler-150",
  "type": "source",
  "process": "main",
  "file_name": "app/main/main-entry.ts",
  "line_num": 150,
  "comment": "Primary open-url event handler that...",
  "source_type": "open-url-handler",
  "source": "app.on(\"open-url\", (event, urlStr) ...",
  "connected_sink": "sink-1-window-mgmt-loadurl-106",
  "introducing_new_taint": true
}, {
  "id": "sink-1-window-mgmt-loadurl-106",
  "type": "sink",
  "process": "main",
  "file_name": ".../window-proc-mgnt-service.ts",
  "line_num": 106,
  "comment": "BrowserWindow.loadURL() called with...", 
  "sink_type": "navigation",
  "sink": "windows.get(id).loadURL(entryURL)",
  "connected_source": "src-1-main-openurl-handler-150",
  "subsequent_source": []
}, ...]
\end{lstlisting}
    
    \begin{lstlisting}[
        language=JavaScript, 
        label={lst:harness-patch},
        caption={
            An example harness patch that injects the fuzzer initialization function in the bootstrap method of the application.    
        }
    ]
    --- a/main.ts
    +++ b/main.ts
    @@ -42,6 +42,7 @@ function bootstrap() {
       initializeWindow();
    +  urlFuzzInitialize();
    }
    \end{lstlisting}
    \vspace{140px}
\end{figure}




\end{document}